\documentclass[preprint,authoryear,12pt,review]{elsarticle}
\usepackage{setspace}
\usepackage{lineno,hyperref}
\usepackage{graphicx}
\usepackage{multirow}
\usepackage{amsmath}
\usepackage{multicol}
\usepackage{algorithm}
\usepackage{ragged2e}
\usepackage{afterpage}
\usepackage{changepage}
\usepackage{geometry}
\usepackage{natbib}
\usepackage{algpseudocode}
\usepackage{array}
\usepackage{booktabs}
\usepackage{makecell}
\usepackage[dvipsnames]{xcolor}
\usepackage{float}
\usepackage[flushleft]{threeparttable}
\usepackage{subcaption}
\usepackage{pdflscape}
\usepackage{dcolumn}
\newcolumntype{d}[1]{D{.}{.}{#1}}
\usepackage{longtable}
\usepackage{capt-of}
\usepackage{lipsum}
\journal{TR part D}
\newcolumntype{L}[1]{>{\raggedright\let\newline\\\arraybackslash}p{#1}}
\newcolumntype{C}[1]{>{\centering\let\newline\\\arraybackslash}p{#1}}
\newcolumntype{R}[1]{>{\raggedleft\let\newline\\\arraybackslash}p{#1}}

\newcommand*\sqcitep[1]{{\setcitestyle{square}$\!\!$\citep{#1}}}

\begin{document}

\begin{frontmatter}

\title{The Role of Urban Form in the Performance of Shared Automated Vehicles}

\author{Kaidi Wang}
\address{School of Public and International Affairs, Virginia Tech}



\ead{kaidi@vt.edu}


\author{Wenwen Zhang\corref{mycorrespondingauthor}}
\cortext[mycorrespondingauthor]{Corresponding Author}
\address{Edward J. Bloustein School of Planning and Public Policy, Rutgers University}
\ead{wenwen.zhang@ejb.rutgers.edu}
\begin{singlespace} 
\begin{abstract}

The technology of Shared Automated Vehicles (SAVs) has advanced significantly in recent years. However, existing SAV studies primarily focus on the system design while limited studies have examined the impacts of exogenous variables, especially urban form, on SAV performance. Therefore, it remains unclear what key urban form measurements may influence SAV system's sustainability. This study fills the research gap by conducting simulation experiments using data collected from 286 cities. This study identifies critical urban form measurements correlated with the simulated SAV performance using {fixed effects} regression models. The results suggest that SAVs are more efficient and generate less VMT in denser cities with more connected networks and diversified land use development patterns. The model results can help provide insights on land use and transportation policies to curb the adverse effects of SAVs in the future and generalize existing SAV simulation results to the rest of U.S. cities.

\end{abstract}

\begin{keyword}
Shared Automated Vehicles (SAVs)\sep Urban Form\sep Simulation\sep Cross-city Comparison\sep SAV Performance
\end{keyword}
\end{singlespace}
\end{frontmatter}

\begin{singlespace}
\section{Introduction}
The commercial deployment of Automated Vehicles (AVs) is almost around the corner. More than 15 companies have announced that they aimed to have automated vehicle technology ready by 2021 \citep{companies2020}. Waymo, an autonomous driving technology company, {\color{Red}has} already started to test commercial deployment of Shared AVs {\color{Red}without safety drivers in Chandler, Arizona \citep{waymo2018}}. {\color{Red}The company has recently opened the self-driving ride-hailing service to the public \citep{waymo2020}.}

The Shared AV (SAV) is a very promising business model of AVs, given its affordability \citep{bansal2016assessing, fagnant2018dynamic, burns2013transforming} and smaller environmental footprint, compared with private AV ownership \citep{wang2019parking, zhang2020parking}. 
On the other hand, compared with conventional vehicles, SAVs may also introduce some undesirable externalities, such as empty or extra Vehicle Miles Traveled (VMT) due to relocation (i.e., SAVs make empty trips to pick up clients) and balancing (i.e., SAVs are pushed to under-served zones) \citep{fagnant_travel_2014, chen2016operations, lu2018multiagent, zhang2017parking}. Some studies also suggest that SAVs may lead to congestion problems \citep{levin2017general, zhao2018anticipating}. Therefore, several Metropolitan Planning Organizations (MPOs) have already recognized the necessity to incorporate AVs in long-range plans to harvest the benefits brought by AVs, while curbing {\color{Red}their} adverse effects \citep{childress2015using, bernardin2019scenario, kim2015travel}. However, {\color{Red}decision-makers face} several challenges {\color{Red}when} incorporating SAVs into the planning process. First, results from existing SAV studies are barely comparable nor generalizable across cities, because simulation experiments are conducted based on different modeling assumptions, simplifications, and model implementation methods \citep{fagnant2018dynamic, zhang2015exploring, chen2016operations, alonso2017demand, lokhandwala2018dynamic}. Second, most existing SAV simulation {\color{Red}efforts} also specifically focus on enhancing the {\color{Red}SAV system design}, such as algorithms associated with vehicle assignment, trip pooling, vehicle relocation, and parking \citep{narayanan2020shared}. {\color{Red}However,} the {\color{Red}impacts} of exogenous variables, {\color{Red}especially} urban form, on SAV performance {\color{Red}remain unclear}. Therefore, there is an urgent need to understand how the performance of SAVs may vary in cities with different urban development and travel patterns and identify key urban form {\color{Red}factors} associated with the sustainability of SAV systems.

To close the aforementioned knowledge gaps, we utilized agent-based simulation and {\color{Red}fixed effects} regression models to examine the role of urban form on the performance of SAVs, using data collected from {\color{Red}286} cities across the U.S. This paper is organized as follows. Section \ref{sec: literature} reviews the existing SAV simulation results {\color{Red}and} key SAV performance metrics to identify research gaps. Section \ref{sec: method} introduces the methodology, including {\color{Red}the design of} SAV simulator, cross-city SAV simulation data collection, and the development of {\color{Red} fixed effects} regression models. Section \ref{sec: results} discusses the results of the simulation and regression models. Section \ref{sec: sensitivity} presents the results of sensitivity tests. Section \ref{sec: conclusion} summarizes the key findings, offers suggestions to planners and policy-makers, and discusses future research potential.

\section{Prior Studies}
\label{sec: literature}
The literature on the {\color{Red} SAV system performance} has proliferated in recent years, as the technology and its commercial deployment potential have advanced significantly. The SAV performance or impact studies have {\color{Red} primarily} focused on enhancing algorithms that can optimize {\color{Red} the system from three aspects, including 1)} the served trips per SAV, {\color{Red} 2)} trip pooling, {\color{Red} and 3)} VMT generation {\color{Red} and}  associated environmental impacts, as summarized in the recently published comprehensive literature review \citep{narayanan2020shared}.

Multiple SAV simulation studies suggest that SAVs hold great potential to reduce vehicle fleet size by serving more trips than conventional vehicles. The number of person trips each SAV can serve per day, however, varies significantly across study areas and model settings, as summarized in Table \ref{tab: review}. {\color{Red} Existing studies suggest one SAV is able to serve from }11 to 980 trips per day. The assumed capacity of SAVs plays a critical role {\color{Red}in} the performance. \cite{alonso2017demand} concluded that a SAV in New York City (NYC) can serve 940 to 980 trips per day if up to 10 people can share a ride while \cite{lokhandwala2018dynamic} claimed that each SAV can serve 60 to 88 trips per day in NYC if only two persons can share a ride. The incorporation of dynamic ride-sharing also influences the model results. SAV simulations with dynamic ride-sharing services tend to report more served {\color{Red}person trips} per SAV per day \citep{chen2016operations,levin2017general,fagnant2018dynamic}. {\color{Red}The results also show} that the number of {\color{Red}person trips} that a SAV can serve on a daily basis is substantially higher in a dense city like NYC \citep{alonso2017demand,lokhandwala2018dynamic,boesch2016autonomous,zhang2016control} compared with more sprawled cities, such as Atlanta \citep{zhang2017parking}, Austin\cite{chen2016operations}, and Orlando \citep{gurumurthy2018analyzing}.

\newgeometry{left=2cm,top=2cm,bottom=2cm,right=2cm}
\begin{landscape}
\small
\begin{longtable}[H]{L{3cm}L{1cm}L{2.5cm}L{2cm}L{2cm}L{2cm}p{9cm}}
\caption{A Comparison of Existing SAV Simulation Results and Model Settings}
\label{tab: review}\\ 
  \toprule
  Author  & Year & Trips/SAV per day & VMT Change & \% pooledTrips & Study Area &  Model Settings\\
  \midrule  
\endfirsthead 
  \toprule
  Author  & Year & Trips/SAV per day & VMT Change & \% pooledTrips & Study Area &  Model Settings\\
  \midrule  
\endhead 
  \midrule 
  \multicolumn{7}{r}{\textit{Continued on next page}} \\ 
  \bottomrule
\endfoot 
  \bottomrule
\endlastfoot 
\citeauthor{fagnant_travel_2014} & 2014 & 35.87 & + 10\% & - & Hypothetical & No dynamic ride-sharing; One new vehicle is generated for clients waiting for more than 10 minutes. \\ 
\citeauthor{martinez2015urban} & 2015 & - & +6\% to +22\% & - &  Lisbon, Portugal & All clients are assumed to be willing to share rides; The detour time and distance should be shorter than 20\% of the original trip; At most 8 passengers can share a vehicle; Detour time is capped at 10 minutes and detour distanec is capped at 2 km; Clients wait for at most 5 minutes.\\ 
\citeauthor{zhang2015exploring} & 2015 & 63.17 & - & 60\% & Hypothetical &  Everyone is willing to share; Two persons can share rides; If both clients can have reduced travel costs, including time and SAV fare costs, then two clients can be pooled together.\\ 
\citeauthor{boesch2016autonomous} & 2016 & 80.7 & - & - & Zurich, Switzerland & No ride-sharing; Clients wait for at most 10 minutes before leaving the system. \\ 
\citeauthor{chen2016operations} & 2016 & 11 to 22 & +7.1\% to 14\% && Austin, TX & No ride-sharing; Maximum wait time is 10 minutes.\\ 
\citeauthor{zhang2016control} & 2016 & 55 & - & - & Manhattan & No ride-sharing; Clients would quit if no empty vehicle is available at the calling station.\\ 
\citeauthor{alonso2017demand} & 2017 & 940 to 980 & - & 90\%+ &  New York City, NY & All clients are assumed to be willing to share rides; High occupancy dynamic ride-sharing; Up to 10 people can share rides.\\ 
\citeauthor{bischoff2017city} & 2017 & - & -15\% to -20\% & - & Berlin, Germany &  All clients are assumed to be willing to share rides; 2-4 persons can share a ride; Maximum total travel time is a portion of direct travel time with ride-sharing; Maximum wait time is 15 minutes.\\ 
\citeauthor{dia2017autonomous} & 2017 & - & +10\% to +29\% & - & Melbourne, Australia & No dynamic ride sharing; Maximum wait time is 0 and 5 minutes\\ 
\citeauthor{jager2017agent} & 2017 & 36.14 & +15.08\% & - & Munich, Germany & No ride sharing.\\ 
\citeauthor{levin2017general} & 2017 & 31.4 (AM 2h) & 0 to +66.67\% & - & Austin, TX & All clients are assumed to be willing to share rides; The capacity of SAVs is 4; Dynamic ride-sharing is based on the portion of extra travel time.\\ 
\citeauthor{zhang2017parking} & 2017 & 32.37 & +33.07\% & - & Atlanta, GA & All clients are willing to share rides; 2 persons can share a ride; Maximum detour is 15\% of direct travel time; Maximum wait time is 15 minutes.\\ 
\citeauthor{fagnant2018dynamic} & 2018 & 32.84 & +4.5\% to +8.7\%& 11\% to 21\%& Austin, TX & Everyone are willing to share; Two persons can share rides; Maximum delay is 20-40\% of travel time without dynamic ride sharing. \\
\citeauthor{gurumurthy2018analyzing} & 2018 & 3 to 32 & -3\% to +1.5\% & - & Orlando, FL & All clients are willing to share rides; At most four persons can share a ride; Maximum wait time is 5 min and 10 min; Maximum extra travel time is 5 min, 10 min, ... 30 min.\\ 
\citeauthor{lokhandwala2018dynamic} & 2018 & 60.63 to 88.18 &  -18\% to -45\%\textsuperscript{2} & 75\%+& New York City, NY & Everyone are willing to share; Two persons can share rides; Maximum delay is set to be a portion of travel time without dynamic ride sharing; The portion follows a triangle distribution, ranging from 0 to 1, with mode set to be 0.5; Clients who are not willing to share ride wait for at most 5 minutes. \\
\citeauthor{system2018reshaping} & 2018 & - &+16\% & - & Boston, MA & No ride-sharing.\\ 
\citeauthor{zhang2020parking} & 2020 & - & +19\% to 176.3\%\textsuperscript{2} & - & Atlanta, GA & No ride-sharing; Maximum wait time is 15 minutes;\\
\end{longtable}
\end{landscape}
\restoregeometry

\end{singlespace}
\begin{singlespace}
A majority of SAV studies also suggest that the system {\color{Red} is} expected to generate extra VMT. Simulation results from studies conducted in cities, such as Lisbon, Portugal \citep{martinez2015urban}, Atlanta, GA \citep{zhang2015exploring}, Austin, TX \citep{chen2016operations, levin2017general, fagnant2018dynamic}, Melbourne, Australia \citep{dia2017autonomous}, Munich, Germany \citep{jager2017agent}, and Boston, MA \citep{system2018reshaping}, show that VMT will increase by a range from 0\% to 176.3\%, as shown in Table \ref{tab: review}. {\color{Red}In contrast,} recent modeling efforts in dense cities, such as New York City (NYC) and Berlin, Germany, suggest that SAVs can reduce VMT generation by 18-45\% \citep{lokhandwala2018dynamic} and 15-20\% \citep{bischoff_autonomous_2016}, correspondingly. The difference in VMT generation results may be attributed to varied model assumptions as well as model environment settings, such as spatial resolution and the design of trip pooling, vehicle assignment, cruising, and parking algorithms. For example, the study in Berlin \citep{bischoff2017city} assumes that ride-sharing clients will tolerate a maximum detour time that is equivalent to the direct travel time, which is longer than most other studies. Meanwhile, NYC and Berlin are substantially denser than the majority of the rest study areas, which may likely result in a higher trip pooling success rate to curb VMT generation. 

The performance of SAVs with dynamic ride-sharing services is typically evaluated using the percentage of successfully pooled trips. The results from existing SAV simulation studies indicate that approximately 11\% to 90\% of the trips that authorized ride-sharing can be successfully pooled together. The design of the trip matching algorithm is one of the key determinants of the trip pooling success rate. For instance, the percentage of successfully pooled trips in NYC can increase from 75\% \citep{lokhandwala2018dynamic}, if only two persons will share rides, to 90\% \citep{alonso2017demand}, if high occupancy dynamic ride-sharing is considered. It is also quite apparent that urban form, especially density, also {\color{Red} tends} to play an important role in influencing the percentage of pooled trips. The overall trip pooling success rate in NYC is 75-90\%, which is significantly higher than the success rate in Austin, TX (i.e., 11\% to 21\% \sqcitep{fagnant2018dynamic}). Meanwhile, the percentage of pooled trips from the simulation conducted under a hypothetical grid-based setting, with national average trip density is approximately 60\%, which is also significantly higher than Austin, TX \citep{zhang2015exploring}. This is because the transportation network in the hypothetical work is grid-based, which makes it much denser and more connected than the network found in the real-world setting. 

The review of recent SAV simulation literature suggests that the simulation results are not consistent and are barely generalizable across cities. Some variations may be explained by different model assumptions and simplifications, as well as the local portfolio of urban form and local traffic flow patterns. This conforms with the consensus achieved in the existing transportation and land use interaction literature \citep{ewing2010travel,friedman1994effect,dunphy1996transportation}, that urban form will influence travel behavior and correspondingly the performance of the transportation system. However, most existing SAV simulation studies are conducted using data from one study area, rendering it hard, if not impossible, to generalize results from one city to another. Furthermore, there is a very limited understanding regarding the impact of urban form on the performance of SAVs to support land use and transportation policymaking in the era of SAVs. {\color{Red} Some studies have conducted sensitivity tests to verify the impact of market penetration \citep{boesch2016autonomous, hyland2020operational}, urban size, and trip density \citep{fagnant_travel_2014} on SAV model outcome under hypothetical urban settings. However, sensitivity tests cannot provide systematic understanding regarding the impact of key urban form variables on SAV performance.} Such a knowledge gap may be attributed to the fact that SAV simulations are data consuming and computationally intensive (especially those with dynamic ride-sharing \sqcitep{martinez2015urban, zhang2015exploring, alonso2017demand, bischoff2017city, levin2017general,zhang2017parking, fagnant2018dynamic, gurumurthy2018analyzing, lokhandwala2018dynamic}). 

Motivated by the above research gap, this study conducts SAV simulation experiments using travel demand data collected from {\color{Red}286 municipalities within 21 metropolitan areas (i.e., outputs of travel demand models built by the Metropolitan Planning Organizations [MPOs]), and develops linear regression models to explore the role of urban form on the performance of the SAV systems, especially service efficiency, VMT generation, and trip pooling.} The knowledge generated by this study can help generalize the existing simulation results to cities where SAV simulation models are not available. 

\section{Methodology}
\label{sec: method}
{\color{Red}This study uses a}gent-based simulation and linear regression models to explore the impact of urban form measurements on the performance of SAV {\color{Red}systems}. We first assess the performance of SAV systems across cities using an agent-based SAV simulation. The associations between SAV system performance and urban form variables, such as density, diversity, and design (i.e., 3D variables) \citep{cervero1997travel} are then quantified using linear regression models. In this section, we first introduce the adopted SAV simulation framework and the estimation of SAV performance metrics based on simulation results. We then describe the collection of travel demand data used to fuel the SAV {\color{Red}simulators} and the criteria for study area selection. Finally, we present the urban form variables preparation method and the development of regression models.

\subsection{Agent-based SAV Simulation and Performance Metrics}
\label{sec: simulation}
\subsubsection{Agent-based SAV Simulator}
{\color{Red}This study builds upon an existing SAV simulator} \citep{zhang2017parking}. The {\color{Red}simulator recreates a centralized dispatching controller that manages} the interactions between travelers and SAV fleets. {\color{Red} The model simulates travelers}  based on (1) Origin-Destination (OD) matrices generated by local travel demand models and (2) the assumed market penetration of SAVs. {\color{Red}The demand module simulates the number of person trips between each pair of origin and destination by generating random numbers that follow Poison distribution with the mean of person trip obtained from the local OD matrices. The departure time for each generated person trip is then simulated based on the trip departure time distribution collected from the 2017 National Household Travel Survey (NHTS).} Whether a traveler is willing to share rides with a stranger is randomly determined based on the willingness to share rate, which is a model assumption controlled by an input parameter. {\color{Red}The system pools two} travelers together if they are both willing to share and the resulting detour time is shorter than 20\% of either trip's direct travel time. The{\color{Red} central controller assigns} SAVs to serve the closest waiting traveler and prioritize{\color{Red}s }sharing over the non-sharing option.
{\color{Red}If no vehicle is available to serve an incoming passenger, the central controller puts the passenger into a waiting list until an SAV is available for assignment. A non-sharing traveler will leave the SAV system after staying in the waiting queue for 10 minutes without being served. Travelers who are willing to share may end up waiting longer than 10 minutes to benefit from splitting SAV costs with other clients. Idle SAVs may cruise to under-served zones (i.e., zones with a supply-demand ratio lower than the city average) to balance the spatial distribution of available vehicles. After cruising to the destination, SAVs will park immediately, if they remain unassigned.} 

We run the model for {\color{Red}two simulation days to} generate results. The first simulation day {\color{Red}is} a warm-up day to obtain the required SAV fleet size. {\color{Red}In the first simulation day, t}he system will keep adding SAVs into the system to fulfill trips generated by clients who have been waiting for 10 minutes. {\color{Red}Multiple prior SAV simulation studies also use this method to determine SAV fleet size \citep{zhang2017parking, fagnant_travel_2014, chen2016operations}.} The simulator will then start to assign a fixed fleet of SAVs (the size is determined by the end of the first simulation day) to fulfill all travel demands in the second simulation day. The simulation incorporates congestion by computing travel time based on loaded networks for morning and evening peak hours. More details regarding the design of the SAV {\color{Red}simulator}, model assumptions, and simplifications can be found in the paper published by \cite{zhang2017parking}. In short, the SAV simulator takes the results of local travel demand models, such as OD matrices and loaded networks by time period, as model inputs and outputs clients' waiting time, required SAV fleet size, and SAV moving trajectories under different levels of market penetration rate and willingness to share. {\color{Red}We then post-process the }SAV moving trajectories to estimate SAV performance, such as average served trips per vehicle,  trip pooling success rate, and VMT generation (both occupied and unoccupied).

In this study, the baseline simulation parameters are set up using prevailing settings in the existing SAV simulation studies, findings from preferences survey, and big data released by ride-hailing companies as listed in Table \ref{tab:param}. The market penetration rate of SAVs is set as one {\color{Red}(i.e., 100\% of trips in the study area will be served by the SAV system)}, an SAV adoption level used in multiple SAV simulation studies \citep{fagnant_travel_2014, bischoff_simulation_2016, gurumurthy2018analyzing, lu2018multiagent, martinez2017assessing}. The clients' level of willingness to share is set as {\color{Red}27.5\%, the average of multiple willingness to share parameters obtained from different sources, as listed in Table \ref{tab:param}.}
{\color{Red}We also conduct s}ensitivity tests in Section \ref{sec: sensitivity} with market penetration levels ranging from 25\% to 100\% and willingness to share levels ranging from 10\% to 50\%  to examine if the impact of urban form on SAV performance tends to vary across scenarios.

\begin{table}[H]
    \centering
    \caption{Baseline Scenario Parameters}
    \label{tab:param}{\color{red}
    \begin{tabular}{p{4cm}p{2cm}p{8cm}}
    \toprule
        Parameter& Value & Parameter settings from prior studies \\ \midrule
        SAV Market Penetration Rate & 100\% & 
        100\% - \citep{fagnant_travel_2014,bischoff_simulation_2016,martinez2017assessing,gurumurthy2018analyzing,lu2018multiagent}\\
        Willingness to Share Rate & 27.5\% & 
        28.24\% - \citep{gilibert2017analysis},\newline
        30\% - \citep{alonso2017demand},\newline 
        25.75\% - calculated by authors using open-sourced ride-hailing trip data from Chicago \citep{noauthor_city_2018}\newline
        30.3\% - calculated by authors using weighted Seattle travel survey \citep{noauthor_household_2017}
         \\\bottomrule
    \end{tabular}
}    
\end{table}

\subsubsection{SAV Performance Evaluation}
\label{subsub: performance}
{\color{Red}We use three} widely used metrics {\color{Red}to assess the performance of SAVs}, namely the number of served trips per SAV, the percent of pooled trips, and the percent of extra VMT. The metrics are calculated using equations \ref{eq: first} to \ref{eq: last}.

\begin{align}
    \label{eq: first}
    Served\ Trips/SAV &= \frac{Number\ of\ served\ trips\ per\ day}{SAV\ fleet\ size} \\
    \%\ of\ pooled\ Trips &= \frac{Number\ of\ successfully\ pooled\ trips}{Number\ of\ trips\ that\ authorized\ ride-sharing} \\
    \label{eq: last}
    \%\ of\ Extra\ VMT &= \frac{Empty\ VMT + Occupied\ VMT - Demanded\ VMT}{Demanded\ VMT}
\end{align}

Where the $Number\ of\ trips\ that\ authorized\ ride-sharing$ is the number of trips generated by clients who are willing to share; the $Empty\ VMT$ incurs when an unoccupied SAV relocates to pick up a waiting client or when {\color{Red}the central controller pushes} an idle SAV from an over-served zone to an under-served zone; the $Occupied\ VMT$ is the VMT generated when at least one passenger is in the SAV; the $Demanded\ VMT$ is the VMT that would have been generated by conventional vehicles if SAVs were not deployed. All the metrics are estimated based on the simulation outputs from the second simulation day, as the first simulation day is set up to determine SAV fleet size and the spatial distribution of SAVs at the beginning of the day.

\subsection{Simulation Data and Study Areas}
\subsubsection{Travel Demand Data Collection}
To fuel the aforementioned simulator, we reached out to MPOs of the top 50 most populous Metropolitan Planning Areas (MPAs) to obtain their latest calibrated travel demand model results. We eventually received travel demand model data from 28 regions and were able to process data from 21 MPOs to conduct simulation experiments. The obtained data include local Traffic Analysis Zone (TAZ)-level OD matrices and loaded network with calibrated link-level travel time for each modeled period, as well as GIS shapefiles of TAZs. {\color{Red}Table \ref{tab: travel demand model} in the Appendix} describes the sources of travel demand data used in this study. For each region, we use the travel demand model results from the baseline model year to fuel the simulators, for two reasons. First, travel demand models are typically calibrated for the baseline year. Second, the assumptions of travel demand forecasts tend to vary across regions and may influence the cross-city comparison. Instead of running simulation at the region level like the travel demand models, we set up separate simulation experiments for selected cities located within these regions to augment the sample size for linear regression model development. {\color{Red}We assume t}he SAV {\color{Red}systems} only serve intra-city vehicle trips{\color{Red}, an assumption }used in the majority of SAV simulation case studies \citep{zhang2017parking, fagnant2018dynamic, gurumurthy2018analyzing, lokhandwala2018dynamic}. The following section discusses the city selection criteria.

\subsubsection{Simulation Boundaries and Study City}
\label{subsub: study area}
Before selecting cities, we first align TAZ and block group outlines with the city administrative boundaries to resolve possible boundary issues. Travel demand models are generally developed at the TAZ level while urban form {\color{Red}is} measured at block group or city level. TAZs, block groups, and city boundaries sometimes are not perfectly aligned with each other. The study results can be biased if the study area for agent-based simulation and urban form measurements are not consistent. Therefore, {\color{Red}we developed} a boundary alignment algorithm automatically align the outlines of TAZs and census block groups with cities' administrative boundaries, as illustrated in Figure \ref{fig: align} {\color{Red}in Appendix}. In other words, the algorithm selects a subset of TAZs and census block groups for each city to match the city administration boundary as much as possible. The algorithm is implemented using Python 3.0 and GeoPandas packages \citep{jordahl2014geopandas}. In some cities, the boundaries of TAZs and census block groups have evolved over time. {\color{Red}We use }TAZ and block group boundaries from the baseline year of the collected travel demand model for boundaries alignment. {\color{Red}Figure \ref{fig: boundary} in the Appendix} illustrates the aligned boundaries of several cities using the proposed algorithm. The results suggest that the designed algorithm can effectively align boundaries, despite some small discrepancies (e.g., south of Minneapolis and west of Atlanta).

\begin{figure}[H]
    \centering
    \includegraphics[width = 0.8\textwidth]{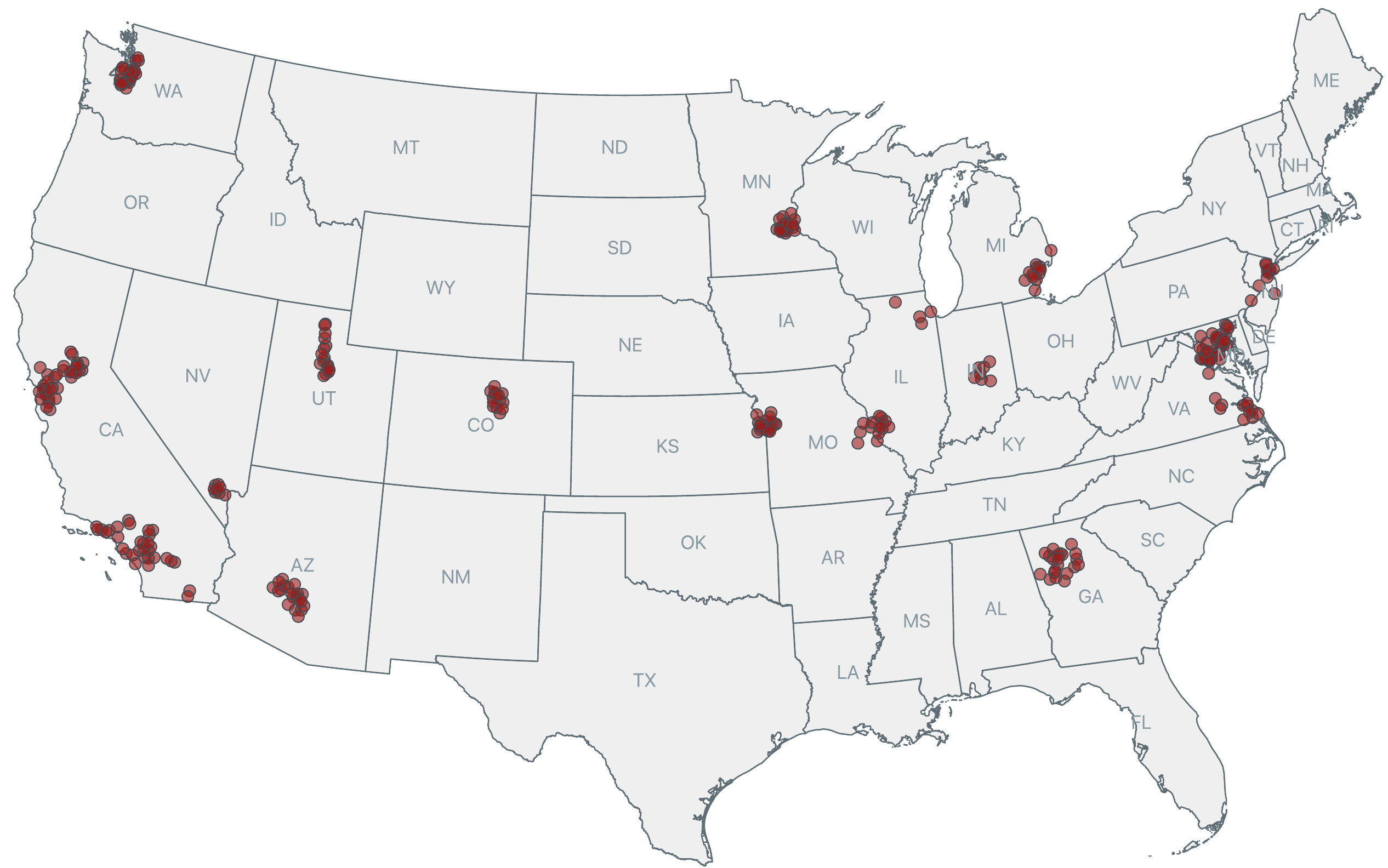}
    \caption{The Locations of Selected Cities}
    \label{fig: locations}
\end{figure}

After aligning the TAZ and block group outlines with city boundaries, we then select cities with more than 10 TAZs and over 20\% of intra-city vehicle trips to conduct simulation experiments. {\color{Red}These city selection criteria are} developed to ensure that the SAVs will serve a decent size of the study area and a reasonable number of vehicle trips. {\color{Red}289 cities meet these selection criteria including three cities with repeated records} (i.e., Bowie, MD, Columbia, MD, and Ellicott City, MD) as they are within the planning boundary of two MPOs (i.e, The Baltimore Regional Transportation Board [BRTB] and The National Capital Region Transportation Planning Board [TPB]). We model these {\color{Red}three} cities using the travel demand model results from TPB, which are calibrated using more recent travel survey data. Finally, {\color{Red}this study simulated 286} cities and their locations are illustrated in Figure \ref{fig: locations}. The cities are quite evenly distributed throughout the States, despite that cities in the middle are slightly undersampled. The selected cities also tend to have a wide range of urban form. There are densely developed cities, such as Los Angelas, CA, Philadelphia, PA, and sparsely built cities, such as Atlanta, GA, and Richmond, VA. To understand if the proposed city selection criteria will influence model results, we also applied the {\color{Red}fixed effect} regression models to several subsets of cities selected using different thresholds of the percentage of intra-city trips for comparison in Section \ref{sec: sensitivity}. 

\subsection{Linear regression model}
\label{sec: regression}
\subsubsection{Independent Variables}
\label{subsub: independent var}
Table \ref{tab: independent var description} summarizes the major categories of independent variables considered in this study. 
Urban form is measured using 3D variables (i.e., density, diversity, and design) \citep{cervero1997travel}. The density variables, examined in this study, include population, household, housing unit, worker, employment densities{\color{Red}, and accumulated accessibility evaluated by job count in each sector}. The population data are collected from 5-year estimates of the American Community Survey (ACS). Workers and employment information are obtained from the Longitudinal Employer-Household Dynamics (LEHD) dataset. The year of ACS and LEHD data {\color{Red}matches} the baseline year of the corresponding city's travel demand model, as listed in Table \ref{tab: travel demand model}. The examined design variables include densities of different types of intersections and networks for automobiles and pedestrians. {\color{Red}Auto-oriented intersections are facilities, where pedestrians are restricted, or the speed is faster than 41 mph, or there are four or more lanes in single direction. Pedestrian-oriented intersections are facilities, where travel speed is lower than 30 mph, or automobile travel is not allowed \citep{epa_2018}.} We accessed these variables from the 2010 Environmental Protection Agency's (EPA) SmartLocation database \citep{epa_2018}. One limitation of using the EPA SmartLocation data is that the baseline years for some models are after 2015, rendering more than five years of {\color{Red}the} time difference between travel demand data and design measurements. However, the changes in the transportation network should be tolerable for model development, especially for city-level design measurements.

Land use diversity is measured using the conventional job-housing entropy, as well as, a graph-based metric, global clustering coefficient (i.e., odCluster). The job-housing entropy is calculated using Equation \ref{eq:entropy}, which is adopted in several prior studies \citep{zhang2018impact, zhang2018residential}. 
\begin{align}
\label{eq:entropy}
jobHouseEntropy &= -\frac{\sum_{j=1}^{5}{p_{ij}\times \log(p_{ij} + 0.01)}}{log(5)}\\
p_{i1} &= \frac{h_{sf}}{h_{sf} + h_{mf} + job_{all}}\\
p_{i2} &= \frac{h_{mf}}{h_{sf} + h_{mf} + job_{all}}\\
p_{i3} &= \frac{job_{reserve}}{h_{sf} + h_{mf} + job_{all}}\\
p_{i4} &= \frac{job_{prof}}{h_{sf} + h_{mf} + job_{all}}\\
p_{i5} &= \frac{job_{labor\ intensive} + job_{resource}}{h_{sf} + h_{mf} + job_{all}}
\end{align}

Where, $h_{sf}$ is single-family housing units, $h_{mf}$ is multi-family housing units; $job_{all}$ is total jobs; $job_{reserve}$ is jobs in retail or service sectors (CNS07, CNS15-20 in WAC table from LEHD data); $job_{prof}$ is jobs in professional sectors, such as finance services (CNS09-13); $job_{labor\ intensive}$ is jobs in labor-intensive sectors, such as manufacturing (CNS03-06, CNS08, CNS14); $job_{resource}$ is jobs in resource-related sectors, such as mining (CNS01-02).

{\color{Red}The job-housing entropy variable captures the diversification of land uses at the zone level. For example, Seattle, WA is more diversely developed, with a median job-house entropy estimated as 0.64, than Kansas City, MO, KS, with a median job-house entropy calculated as 0.50.} 

\newgeometry{left=2cm, bottom=2cm, right=2cm}

    \begin{landscape}
        \small
        \begin{longtable}[h]{p{4.5cm}p{3.5cm}p{9cm}p{2cm}p{2cm}}
\caption{Summary of Independent Variables}
\label{tab: independent var description}\\ 
  \toprule
  Variable & Data Source & Description (unit) & Geographical Units & Aggregation Method\\
  \midrule  
\endfirsthead 
  \toprule
  Variable & Data Source & Description (unit) & Geographical Level & Aggregation Method\\
  \midrule  
\endhead 
  \midrule 
  \multicolumn{5}{r}{\textit{Continued on next page}} \\ 
  \bottomrule
\endfoot 
  \bottomrule
\endlastfoot 
\textbf{Density} &&&&\\
popDen & ACS5 & Population density (per square mile) & City & N/A\\ 
hhDen & ACS5 & Household density (per square mile) & City & N/A \\ 
hsDen & ACS5 & House density (per square mile) & City & N/A \\ 
workerDen & LEHD & Worker density (per square mile) & City & N/A\\ 
jobDen & LEHD & Job density (per square mile) & City & N/A\\
{\color{Red}jobService} & {\color{Red}LEHD} & {\color{Red}Service job count within [1,3,5,10,20,40,60,80] miles of the centroid of block groups} & {\color{Red}Block group} & {\color{Red}Median and mean}\\
{\color{Red}jobRetail} & {\color{Red}LEHD} & {\color{Red}Retail job count within [1,3,5,10,20,40,60,80] miles of the centroid of block groups} & {\color{Red}Block group} & {\color{Red}Median and mean}\\
{\color{Red}jobOffice} & {\color{Red}LEHD} & {\color{Red}Office job count within [1,3,5,10,20,40,60,80] miles of the centroid of block groups} & {\color{Red}Block group} & {\color{Red}Median and mean}\\
{\color{Red}jobIndust} &{\color{Red} LEHD} & {\color{Red}Industry job count within [1,3,5,10,20,40,60,80] miles of the centroid of block groups} & {\color{Red}Block group} & {\color{Red}Median and mean}\\
{\color{Red}jobEntertain} & {\color{Red}LEHD} & {\color{Red}Entertainment job count within [1,3,5,10,20,40,60,80] miles of the centroid of block groups} & {\color{Red}Block group} & {\color{Red}Median and mean}\\
\hline
\textbf{Design} &&&&\\
intersect3DenPed & SmartLocation & Pedestrain-oriented 3-leg intersection density (per square mile) & City & N/A \\ 
intersect4DenPed & SmartLocation & Pedestrain-oriented 4-leg intersection density (per square mile) & City & N/A \\ 
intersectDenNonAuto & SmartLocation & Non-auto oriented intersection density (per square mile) & City & N/A\\ 
intersectDenAuto & SmartLocation & Auto oriented intersection density (per square mile) & City & N/A \\ 
netDenPed & SmartLocation & Pedestrain-oriented network density (miles of links per square mile) & City & N/A\\ 
netDenAuto & SmartLocation & Auto-oriented network density (miles of links per square mile)  & City & N/A\\
\hline
\textbf{Diversity} &&&& \\
odCluster & LEHD & Global clustering coefficient for the graph of commuter flow & City & N/A \\ 
jobHouseEntropy & ACS5 + LEHD & Job house entropy \citep{zhang2018residential} & Block group & Median and mean\\ 
\hline
\textbf{Control} &&&& \\
landSqml & Census Geography & Land area in square mile & City & N/A \\
speed & Travel Demand Model & Average link-level speed in AM peak & City & Median and mean\\
\end{longtable}
\end{landscape}

\restoregeometry
\end{singlespace}
\begin{singlespace}
The global clustering coefficient measures the degree to which nodes (i.e., home and work block groups for this study) in a weighted graph tend to cluster together \citep{opsahl2009clustering}. {\color{Red} Multiple recent urban studies use the variable to measure the morphology of street networks \citep{jiang2014topological} and human activity/travel demand patterns \citep{li2017measuring,he2020demarcating,saberi2017complex}. The results show that the travel demand weighted global clustering coefficient can effectively measure urban centrality \citep{saberi2017complex}. Cities with strong employment centers (i.e., higher urban centrality) tend to have more segregated land use structures, which may significantly increase SAV empty VMT generation. This study introduced travel demand weighted global clustering coefficient into the model because convectional land use diversity measures and accessibility indices cannot effectively capture urban centrality.} In this study, we imputed the global clustering coefficient based on the Graph with links weighted using commuting flows obtained from LEHD Origin-Destination Employment Statistics (LODES) data. 
The coefficient is calculated using built-in functions from the Python NetworkX package \citep{hagberg2008exploring}.

In addition to 3D variables, we also included several control variables, including land area and link-level speed during AM peak hours, as prior SAV simulation studies suggest these variables may influence the performance of SAVs \citep{fagnant_travel_2014, zhang2017parking, chen2016operations}. AM peak hours are defined by the local travel demand models, which vary slightly across regions. {\color{Red}We also introduced the region dummy variables to control the fixed effects that regions may have on the performance of SAV systems, stemming from the assumptions and simplifications of local travel demand model.}

{\color{Red}There is a total of 48 independent variables in this study including 21 region dummy variables. Table \ref{tab: independent var description} lists the initial pool of independent variables. Specifically, we used the median and mean values of job count in each sector within specific distances, job-house entropy, and link-level speed to aggregate block group level measurements to city level.}

\subsubsection{Regression Model Development}
 We used multiple linear regression models to quantify the correlations between urban form variables and SAV performance. {\color{Red}We log-transformed skewed performance variables and independent variables (i.e., variables whose absolute values of kurtosis are larger than three).} Urban form variables are highly correlated with each other, which will lead to multicollinearity problems. Some prior studies used Principal Component Analysis (PCA) method to transform urban form variables to avoid such an issue \citep{ewing2002measuring}. However, the PCA approach may result in model interpretation issues. Therefore, we applied a customized forward stepwise regression algorithm to select independent variables for the final model. Previous modeling efforts also used such modeling approach for models involving a large number of built environment variables \citep{le2018correlates, hankey2017population}. First, {\color{Red}the stepwise feature selection algorithm introduces} the variable with the highest correlation with the dependent variable into the model. {\color{Red}The algorithm} then iteratively {\color{Red}adds} the independent variable that is most correlated with the model residuals and is not highly correlated with the included independent variables (i.e., Pearson correlation $>0.75$). In each iteration, the algorithm re-estimates the linear model with the newly added independent variable to determine if the introduced independent variable violates the rule of multicollinearity (i.e., VIF $>5$) and should be removed. This process stops if the last added independent variable does not increase the adjusted $R^2$ of the model or the estimated coefficient for the last introduced variable is not significantly different from zero. {\color{Red}Finally, we applied Moran's I tests on regression model residuals to determine if there are spatial autocorrelations. We defined the city pair-wise weights as the inverse of the distance between their centroids.}


\section{Results and Discussion}
\label{sec: results}
\subsection{Simulation Model Results}
\label{sec: sim results}

\begin{figure}[h]
    \begin{subfigure}{0.49\linewidth}
    \includegraphics[width = \linewidth]{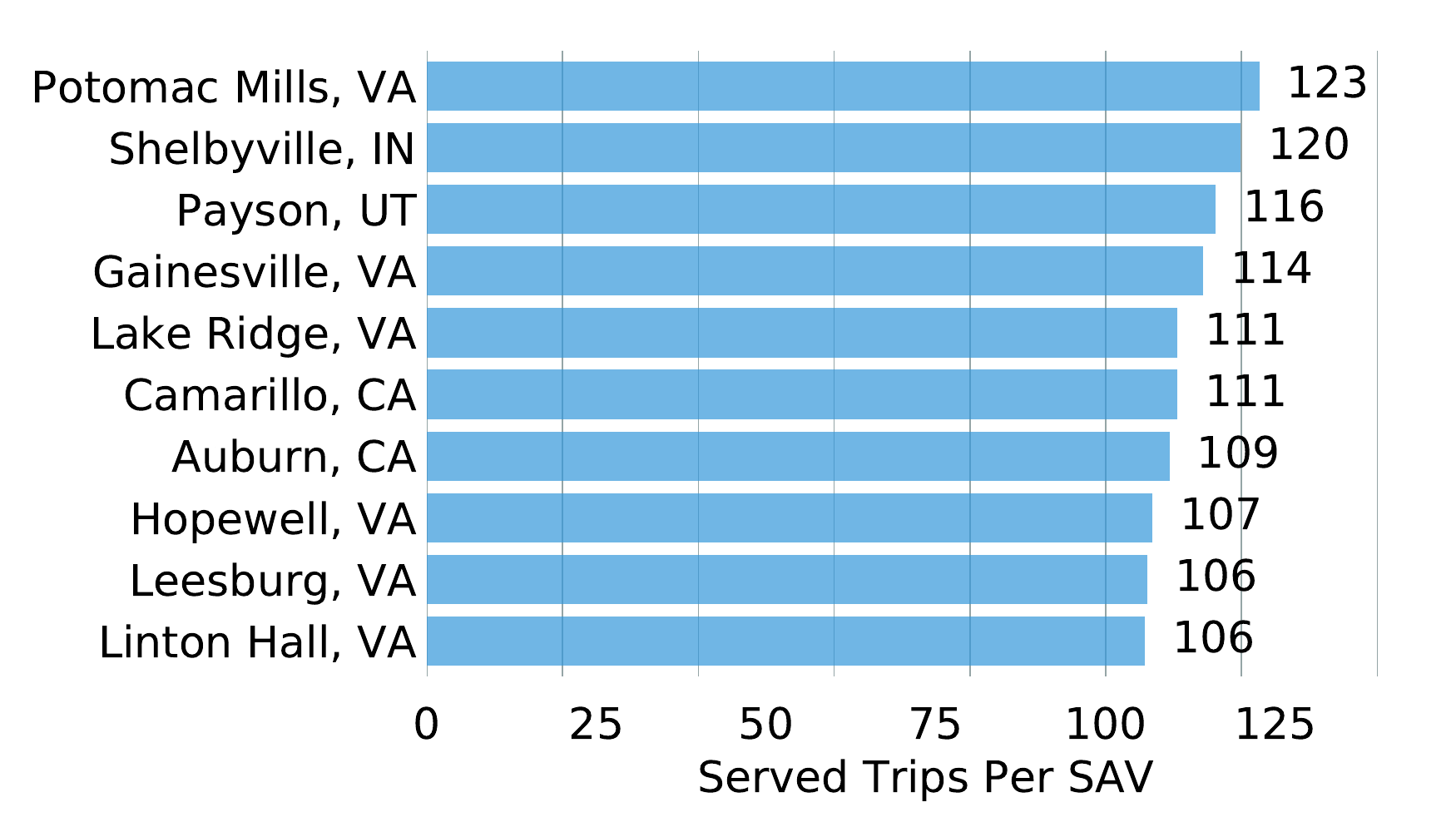}
    \caption{Top 10 Cities by Served Trips Per SAV}
    \end{subfigure}
    \begin{subfigure}{0.49\linewidth}
    \includegraphics[width = \linewidth]{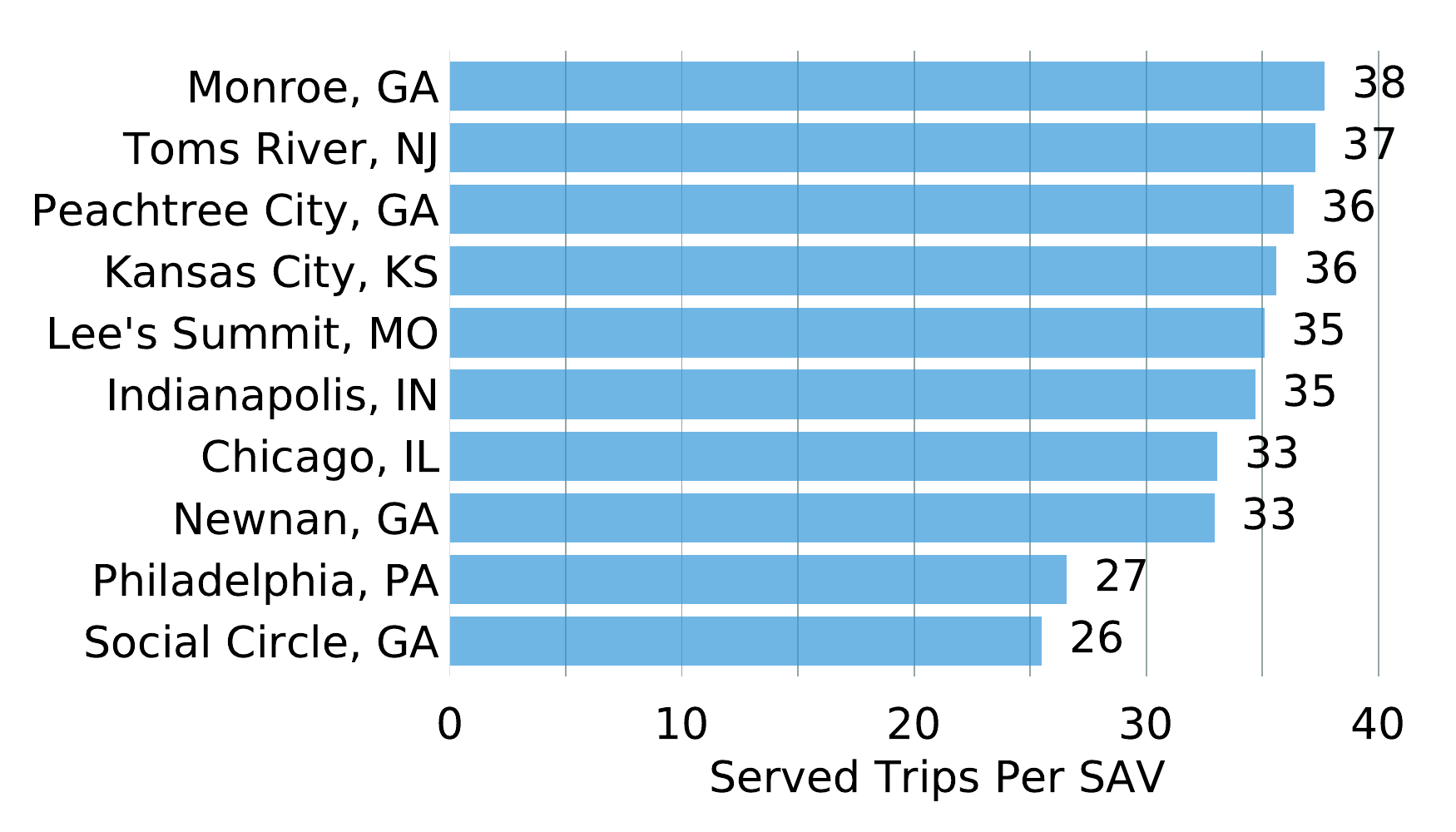}
    \caption{Bottom 10 Cities by Served Trips Per SAV}
    \end{subfigure}
    \begin{subfigure}{0.49\linewidth}
    \includegraphics[width = \linewidth]{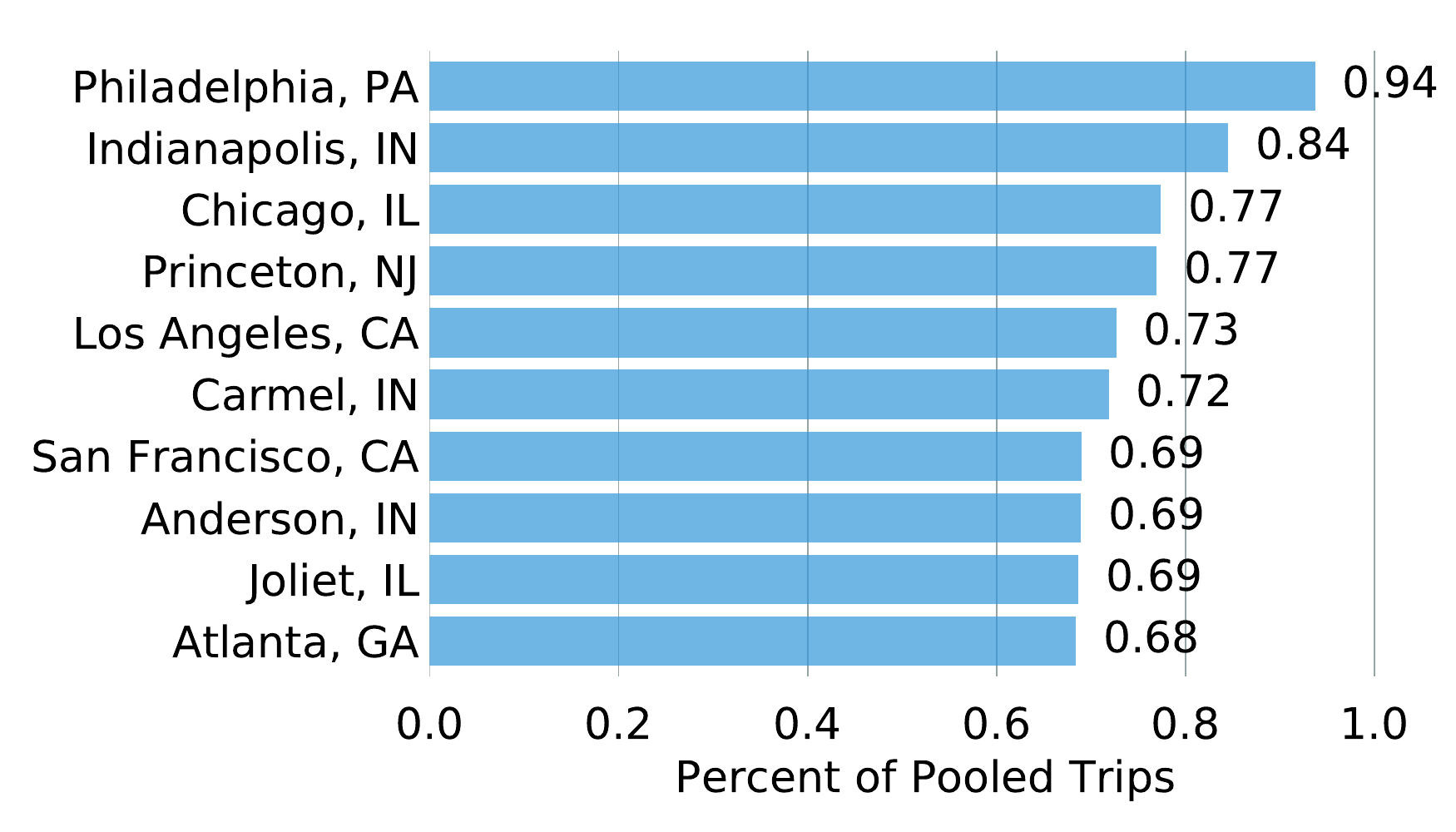}
    \caption{Top 10 Cities by Percent of Pooled Trips}
    \label{fig: top pool}
    \end{subfigure}
    \hfill
    \begin{subfigure}{0.49\linewidth}
    \includegraphics[width = \linewidth]{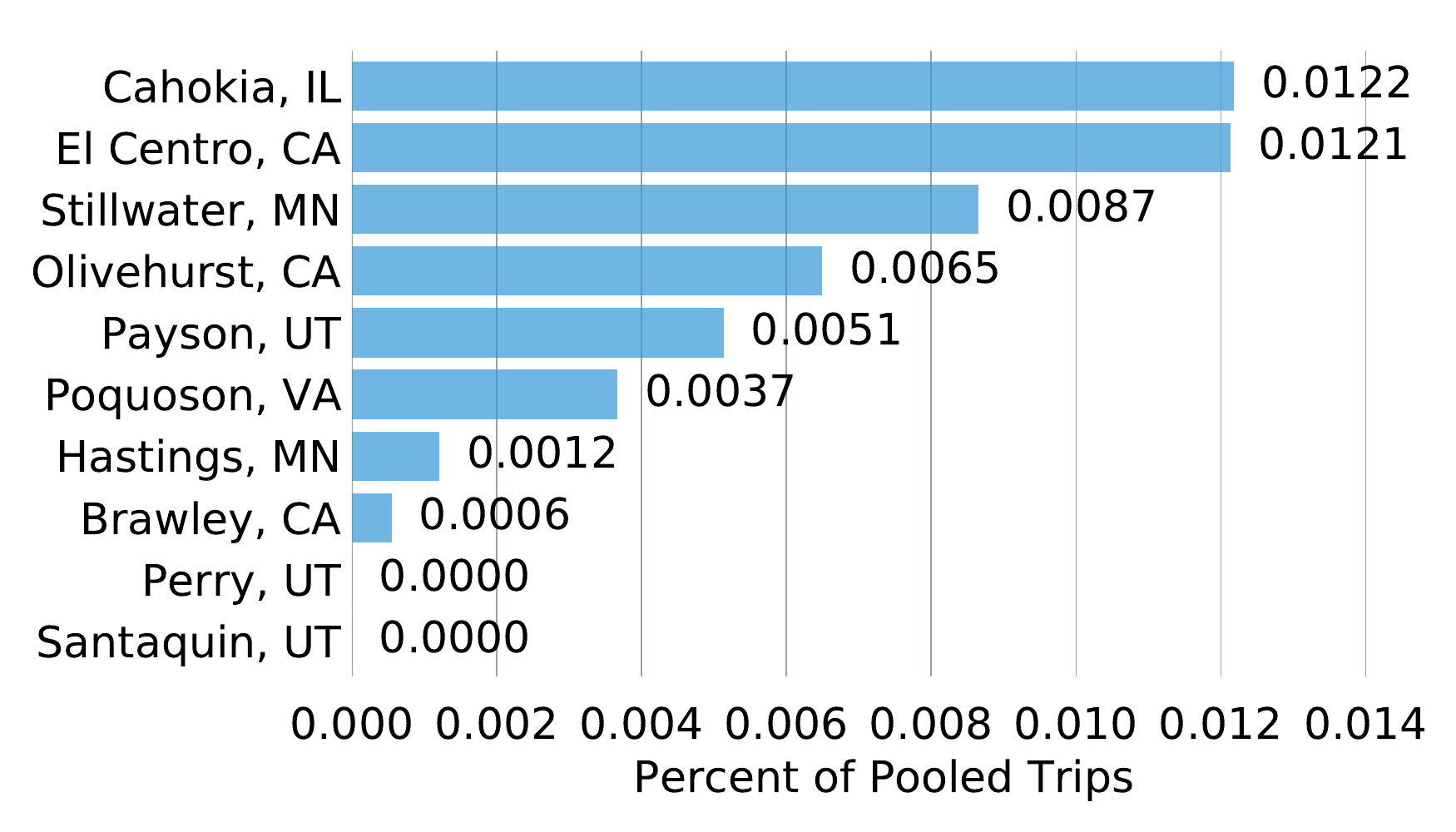}
    \caption{Bottom 10 Cities by Percent of Pooled Trips}
    \label{fig: bottom pool}
    \end{subfigure}
    \begin{subfigure}{0.49\linewidth}
    \includegraphics[width = \linewidth]{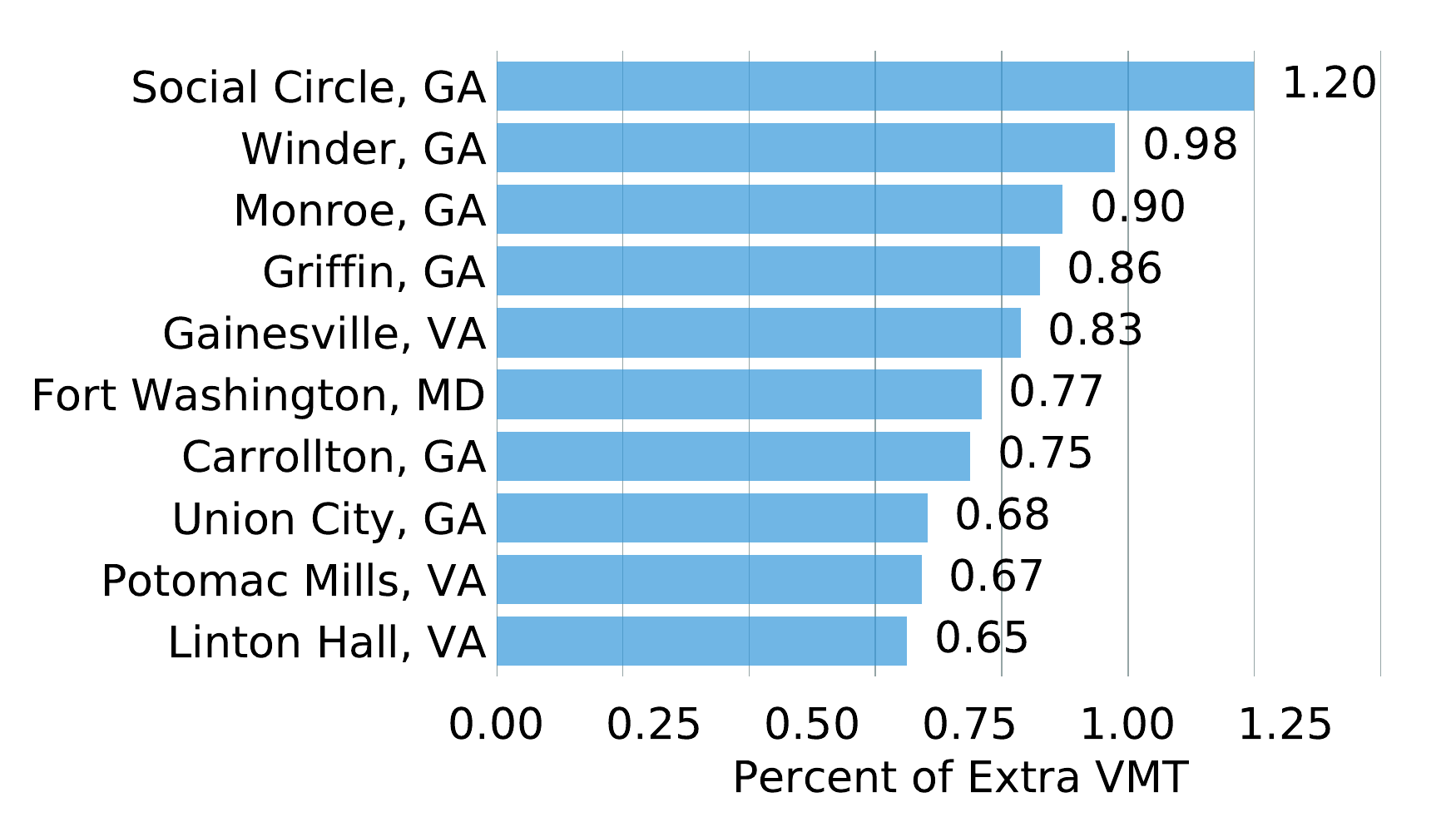}
    \caption{Top 10 Cities by Percent of Extra VMT}
    \end{subfigure}
    \hfill
    \begin{subfigure}{0.49\linewidth}
    \includegraphics[width = \linewidth]{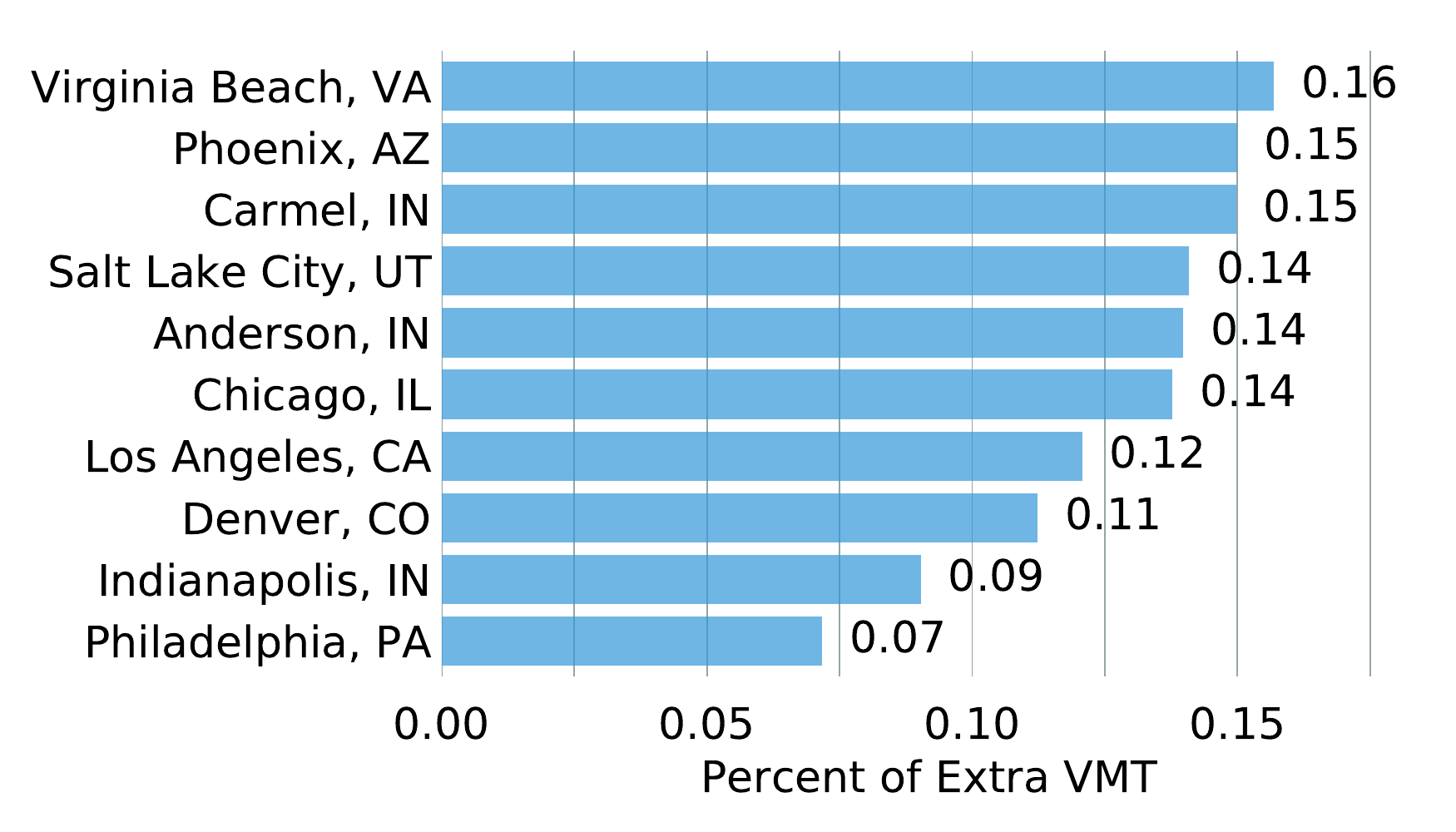}
    \caption{Bottom 10 Cities by Percent of Extra VMT}
    \label{fig: bottom extra}
    \end{subfigure}
    \caption{Top and Bottom 10 Cities by SAV Performance}
    \label{fig: top_bottom}
\end{figure}

Figure \ref{fig: top_bottom} shows the top and bottom ten cities by SAV performance metrics. {\color{Red}The supplementary document provides a table} of model results for reference. Additionally, we also open sourced the simulation results on \href{https://github.com/nacici/Urban-form-and-SAV-performance}{\color{Blue}a Github repository} for easy access. The results suggest SAV performance can vary dramatically across cities, indicating that it is critical to examine the exogenous determinants of such variation to obtain more generalizable knowledge regarding the possible impacts of SAVs. Specifically, the simulated average number of trips served by each SAV ranges between 26 (in Social Circle, GA) and 123 (in Potomac Mills, VA). {\color{Red}The percent of successfully pooled trips also varies dramatically across cities given the same willingness to share rate. No trips are pooled in Perry, UT and Santaquin, UT while 93.72\% of simulated person trips that are willing to share rides are pooled as shown in Figure \ref{fig: top pool} and \ref{fig: bottom pool}. Meanwhile, the substantial variation of the percent of extra VMT suggests that SAVs may operate efficiently in some cities, such as Philadelphia, PA, while leading to congestion and environmental issues in cities like Social Circle, GA.}

\newgeometry{top=2cm}
\restoregeometry
\end{singlespace}
\begin{singlespace}
\subsection{Regression Model Results}
Table \ref{tab: regression_estimate} shows the estimated coefficients for significant variables (i.e., $p <0.05$) by SAV performance metrics. {\color{Red}Table \ref{tab:dummy} in Appendix illustrates the model estimates on the region dummy variables. The estimated P-values for Moran's I test are 0.127, 0.333, and 0.122 for the model of trips per vehicle, percent of pooled trips, and the percent of extra VMT correspondingly. This indicates there is no spatial autocorrelation at 95\% significant level, after controlling for regional fixed effects.}

\begin{table}[H]
    \centering
    \hspace*{-1cm}
    \begin{threeparttable}
        \caption{Model Estimate by Performance Metrics}
        \label{tab: regression_estimate}\color{Red}
        \begin{tabular}{p{3.5cm}ccc}
        \toprule
        Variable & ServedTrips/SAV & \%pooledTrips & Log(\%extraVMT)\\
        \midrule
             jobDenLog & - & 0.089\textsuperscript{***} & -\\
             netDenPed & 1.580\textsuperscript{**} & - & - \\
             intersectDenNonAutoLog & - & - & -0.331\textsuperscript{***} \\
             jobHouseEntropyMedian & - & - & -0.028\textsuperscript{**} \\
             odClusterLog & - & -0.023\textsuperscript{**} & 0.039\textsuperscript{**} \\
             landSqmlLog & -11.219\textsuperscript{***} & 0.115\textsuperscript{***} & -0.271\textsuperscript{***}\\
             speedAveLog & 12.953\textsuperscript{**} & - & - \\
         \midrule
         N & 286 & 286 & 286\\
         Adjusted $R^2$& 0.812 & 0.880 & 0.854\\
         P value of Moran's I test & 0.127 & 0.333 & 0.122 \\
        \bottomrule
        \end{tabular}
        \begin{tablenotes}
          \small
          \item
          \textsuperscript{***}$p<0.001$, \textsuperscript{**}$p<0.01$, \textsuperscript{*}$p<0.05$.
        \end{tablenotes}
    \end{threeparttable}
    \hspace*{-1cm}
\end{table}

\subsubsection{Served Trips Per SAV}
The simulated number of served trips per SAV is primarily associated with urban design. The model explains {\color{Red}81.2}\% of variation in served trips per SAV. Pedestrian-oriented {\color{Red}network} density is positively associated with the served trips per SAV. One {\color{Red}standard deviation} increase in pedestrian-oriented {\color{Red}network} density is associated with {\color{Red}1.58} more trips served per SAV per day. Cities that are more pedestrian-oriented tend to have smaller blocks and more connected transportation networks, which may result in shorter network distances between origins and destinations.


{\color{Red} The congestion level and land area (i.e., SAV operation area size)} can also influence the number of trips each SAV can serve per day. SAVs can serve more trips in cities with a lower congestion level while holding other variables constant. For a 1\% increase in the average link-level travel speed during AM peak hours, we expect each SAV to serve around {\color{Red}0.12} more trips per day. {\color{Red}Fewer trips can be served per SAV if the land area is larger, as the trips tend to be longer in larger SAV service areas}. On average, a 1\% increase in service area is expected to reduce the number of served trips per SAV per day by {\color{Red}0.11. This finding is consistent with prior SAV simulation results \citep{fagnant_travel_2014}}.

The results suggest that {\color{Red}improving network facilities for pedestrians and reducing urban sprawl may enhance SAV efficiency. Meanwhile, transportation and land use policies that can reduce congestion may also increase SAV efficiency. For instance, promoting alternative mode travel through investing in pedestrians oriented infrastructure designs may improve SAV performance in the city.}

\subsubsection{Percent of Pooled Trips}
\label{subsub: percent pool}
Density and diversity factors have an influence on the percent of pooled trips after controlling for the level of willingness to share. The adjusted $R^2$ for the model is {\color{Red}0.880}, indicating the majority of the variation in the percentage of pooled trips is explained by the model. Holding other variables constant, cities with higher job density tend to have more trips successfully pooled together. One percent increase in job density will increase the percentage of pooled trips by {\color{Red}0.089} percentage points. This is quite expected, as trips with closer origins and destinations in dense cities are more likely to be pooled together.


Land use diversity {\color{Red}at city level} may also enhance the success rate of trip pooling. The {\color{Red}travel demand weighted global clustering coefficient is negatively correlated with the percentage of pooled trips. One percent increase in the travel demand weighted global clustering coefficient may reduce trip pooling success rate by 0.023} percentage points. The higher the global clustering coefficient, the more segregated and monocentric the city is. Therefore, polycentric cities with mixed land use development patterns may have more trips pooled together. 


{\color{Red}Land area has a positive effect on the percentage of pooled trips. One percent increase in the service area will increase the ride-sharing success rate by 0.115 percentage points. This may be attributed to the design of the trip pooling algorithm in the SAV simulation model. Two trips can be pooled together when each client's detour time is shorter than 20\% of their direct travel time. Travelers in larger cities tend to make longer trips and tolerate longer detours. As a result, more trips can be pooled in larger cities.}

In sum, the results show that cities may improve the percent of pooled trips by promoting compact development and enhancing land use diversity.

\subsubsection{Percent of Extra VMT}
The model presented in the study can explain {\color{Red}85.4\%} of the variation in the percent of extra VMT. Non-auto-oriented intersection density is negatively correlated with the percentage of extra VMT. One percent increase in the intersection density is related to a {\color{Red}0.331}\% decline in the percent of extra VMT. {\color{Red} This may be because a more connected network can reduce  vehicle relocation distance for client pick up or curb detour distance for ride-sharing trips}.

Diversity is also negatively associated with extra VMT generation. One {\color{Red}deviation} increase in {\color{Red}median job-house entropy} is associated with {\color{Red}2.84\% (exp[0.028] - 1)} decline in the percent of extra VMT. We expect the percentage of extra VMT to be increased by {\color{Red}0.039\%} for every {\color{Red}percent} increase in the commuting flow global clustering coefficient. This is probably because diversity can reduce empty VMT stemming from empty cruising and pick up trips. {\color{Red} Additionally, polycentric cities (with lower travel demand weighted global clustering coefficient) can reduce extra VMT generation due to the higher trip pooling success rate}, as discussed in the last subsection.

{\color{Red}Land area has a negative effect on the percent of extra VMT. One percent increase in land area can reduce the percent of extra VMT by 0.271\%. As we mentioned in the model of trip pooling success rate, trips tend to be longer in larger cities and longer trips can enhance trip pooling success rate. Therefore, the share of extra VMT tends to be smaller in larger places as ride-sharing can help curb the VMT generation.}




The regression results suggest planners and decision-makers may consider using transportation and land use policies to increase network connectivity by providing more facilities for non-auto modes, as well as encourage diversified land use development patterns to reduce extra VMT generation. 

\section{Sensitivity Tests}
\label{sec: sensitivity}

\subsection{Threshold of the Percentage of Intra-city Trips}
In this study, we develop regression models using SAV simulation results for cities with more than 20\% intra-city trips. To determine if such city selection criteria will influence the model output, we compare the presented results with regression model results estimated using cities selected with different intra-city trip selection thresholds. Root mean square errors (RMSE) of the models are utilized to evaluate the performance of the models on cities selected with various thresholds. We also refit the models on new subsets of cities and compared the coefficients across different thresholds with t-tests. Specifically, we test samples selected with the thresholds of the percentage of intra-city trips ranging from 20\% to 45\%. 

\begin{figure}[H]
    \centering
    \includegraphics[width = 0.8\linewidth]{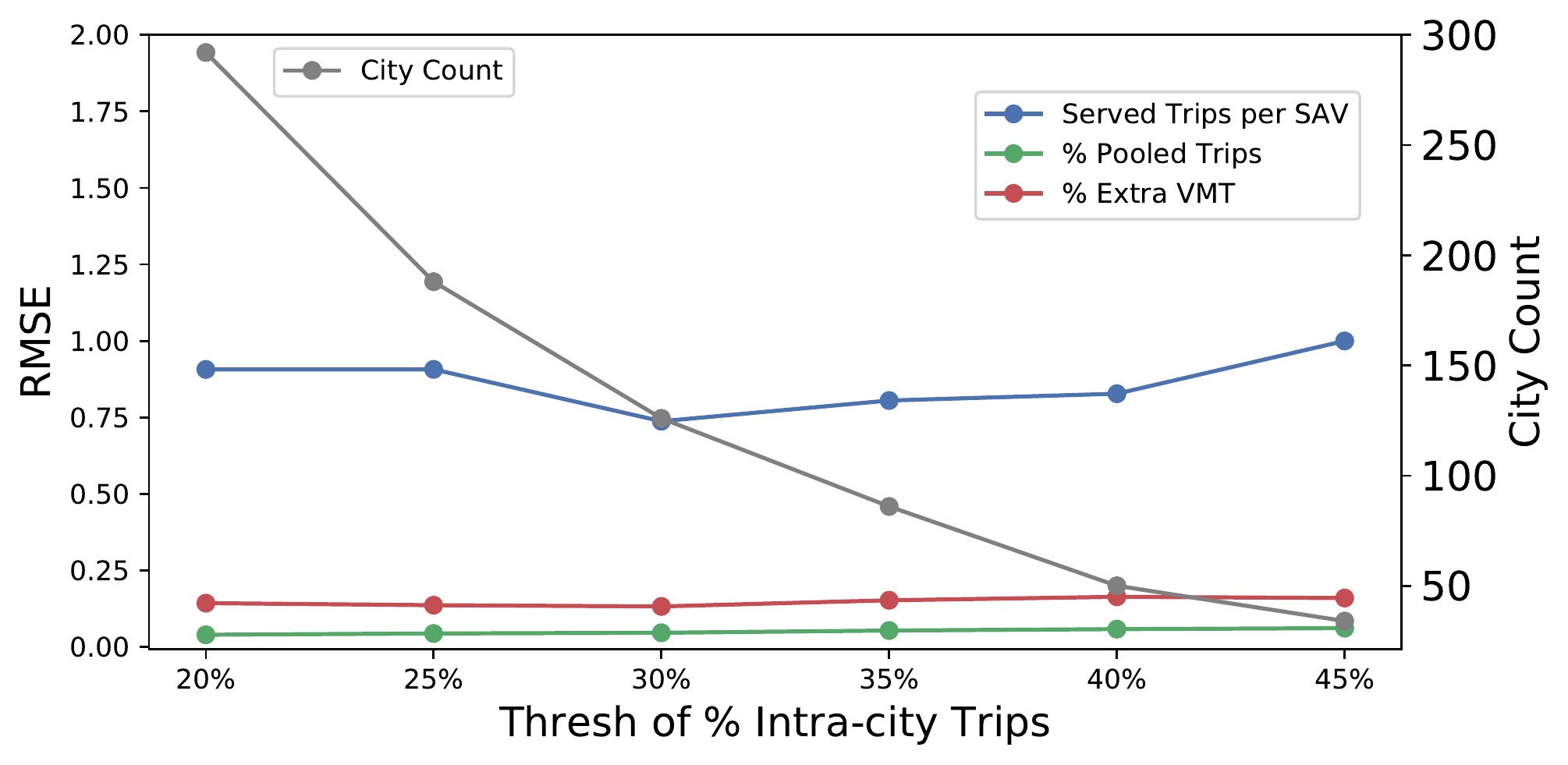}
    \caption{\color{Red}RMSE When the Fitted Models are Applied to Various Samples}
    \label{fig: thresh RMSE}
\end{figure}

\begin{figure}[H]
    \centering
    \includegraphics[width = \linewidth]{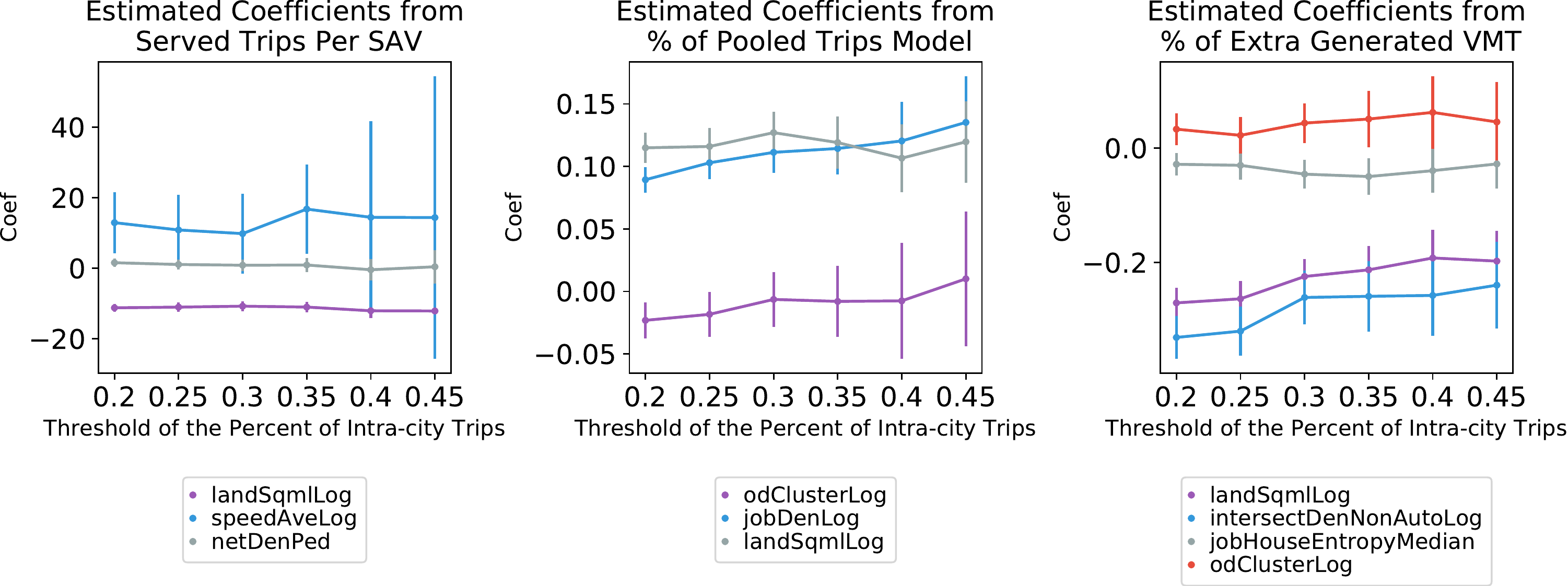}
    \caption{\color{Red}Estimated Coefficients and 95\% Confidence Interval of Each Model Using Various Thresholds of Percentage of Intra-city Trips}
    \label{fig: thresh coef}
\end{figure}

Figure \ref{fig: thresh RMSE} illustrates the RMSEs of linear regression models with different thresholds for the percentage of intra-city trips. We rescaled the RMSEs for the served trips per SAV to 0 to 1 for the comparison purpose. RMSEs for the percent of pooled trips and the percent of extra VMT models are almost constant across scenarios. The RMSEs of the served trips per SAV model change slightly. This may be attributed to the fact that a significantly smaller number of cities are included in the model when the threshold increases. The small sample sizes (especially in 40\% and 45\% scenarios) may not be representative and lead to inaccurate estimates of the RMSEs. Figure \ref{fig: thresh coef} shows the estimated coefficients and their 95\% confidence intervals by SAV performance metrics across various thresholds. The coefficients only change slightly across different threshold scenarios. The differences in estimated coefficients are not significantly significant based on the results of t-tests. Therefore, we argue that the city selection method used in this study will not significantly influence model outputs.

\subsection{Willingness to Share Rides and Market Penetration Rate}
\label{sec: model sensitivity}
The SAV performance variables (i.e., dependent variables) are simulated based on the assumptions that the willingness to share ride rate (WS) is 27.5\% and the market penetration rate of SAV (MP) is 100\%. We conduct sensitivity tests to evaluate how changes in WS and MP may influence the role of urban form. First, for each selected city, we conduct more simulation experiments with the willingness to share rides ranging from 10\% to 50\% and the market penetration rate ranging from 25\% to 75\%. Second, we re-estimate regression models, using the SAV performance metrics generated by the new simulation experiments. Finally, we conduct t-tests to determine if the differences in the estimated coefficients across scenarios are statistically significant.

\begin{figure}[H]
    \centering
    \includegraphics[width =\linewidth]{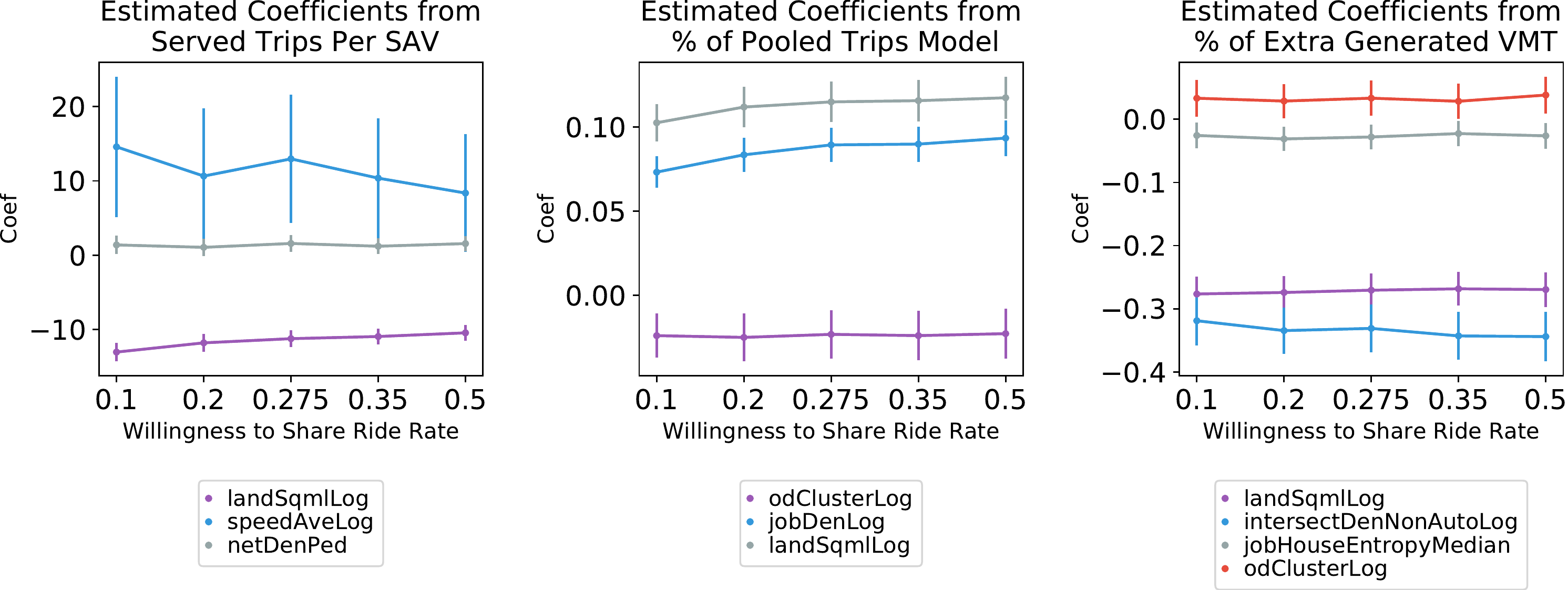}
    \caption{\color{Red}Coefficients with Various Willingness to Share}
    \label{fig: sensitivity_ws}
\end{figure}

\begin{figure}[H]
    \centering
    \includegraphics[width =\linewidth]{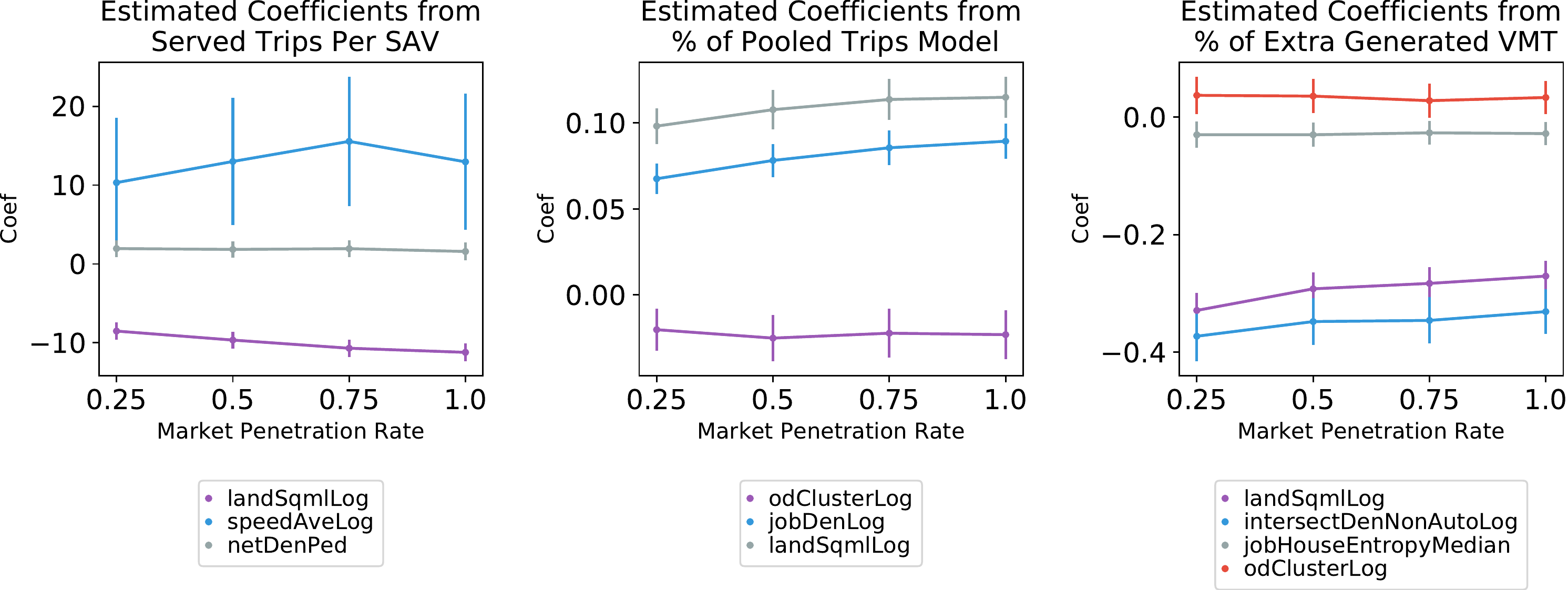}
    \caption{\color{Red}Coefficients with Various Market Penetration rates}
    \label{fig: sensitivity_mp}
\end{figure}

Figure \ref{fig: sensitivity_ws} and \ref{fig: sensitivity_mp} show the estimated coefficients of independent variables and the 95\% confidence intervals by SAV performance metrics across different WS and MP scenarios. The results suggest that the variation in the estimated coefficients is marginal, as the plotted curves are quite flat. The t-test results suggest that the differences are not significant. Therefore, our results are robust across the examined SAV deployment scenarios. 

\section{Conclusion}
\label{sec: conclusion}
This study contributes to the SAV literature by providing insights into the impact of urban form on the performance of SAV systems. Specifically, the examined SAV performance metrics include served trips per SAV per day, the percent of pooled trips, and the percent of extra VMT. These performance metrics are generated using an SAV simulator \citep{zhang2017parking} using travel demand model data collected from 286 cities located within the service boundary of 21 MPOs, assuming all vehicle trips are served by SAVs and 27.5\% of travelers are willing to share rides. Fixed effects linear regression models are then developed to identify the key urban form measurements (i.e., Density, Diversity, and Design variables) associated with the simulated SAV performance metrics.

{\color{Red}The modeling results suggest that urban form can significantly influence the performance of SAVs. Specifically, the pedestrian-oriented network density and intersection density can reduce extra VMT generation and improve SAV efficiency. Besides, policies that encourage more compact employment distributions may effectively increase trip pooling success rates. {\color{Red}Land use policies increasing land use diversity (i.e., higher job-house entropy)} and supporting multi-core development (i.e., lower travel demand weighted global clustering coefficient) may also promote SAV performance. {\color{Red}The results also suggest that local congestion levels may have a negative impact on the number of person trips served per SAV. Therefore, land use policies that encourage compact development may be bundled with travel demand management strategies, such as alternative travel mode promotion policies that mitigate local congestion, to enhance the efficiency of SAVs.}
}

The sensitivity test results justify our city selection method and SAV simulation model assumptions. The linear model results are quite constant when selecting cities with 20\%-45\% of intra-city trips. Furthermore, model results across scenarios with the willingness to share ranging from 10\% to 50\% and market penetration ranging from 25\% to 100\% are not significantly different based on t-test results, indicating the influence of the urban form on SAV performance is not sensitive to changes in these two model assumptions.

There are some limitations that merit future research efforts. In this study, we only examined three SAV performance metrics. Future studies can extend knowledge by exploring more performance metrics, such as parking demand and energy and emissions. We did not include parking demand analysis in this study due to a lack of local parking demand data incorporating parking prices. The impact of urban form on SAV energy and emissions are not analyzed, as we assumed that SAVs would only serve existing vehicle trips, and no modal shift is considered. Future studies may relax such model assumptions to gain more insights on how urban form can alleviate or exaggerate the negative environmental externalities of SAVs. {\color{Red}Finally, the model also adopts the assumption of homogeneous willingness to share and market penetration rate of SAVs, while urban form may also influence these factors. Prior studies suggest that preference for SAV adoption can be influenced by urban form, built environment, and transit infrastructures \citep{bansal2016assessing, hajjafari2018exploring, nazari2018shared}. One recent study has also developed a machine-learning based microsimualtion approach to integrate neighborhood-level SAV preference into the agent-based simulation \citep{zhang2020synthesizing}. Future research can extend this study by incorporating heterogeneous SAV and ride-sharing preferences}. 


\section*{Acknowledgements}
We thank the MPOs who provide us with the travel demand model output. These MPOs include Hampton Roads Transportation Planning Organization (HRTPO), Metropolitan Transportation Commission (MTC), National Capital Region Transportation Planning Board (TPB), Wasatch Front Regional Council, Metropolitan Council, Chicago Metropolitan Agency for Planning (CMAP), Maricopa Association of Governments (MAG), Regional Transportation Commission of Southern Nevada (RTC), Puget Sound Regional Council (PSRC), Southeast Michigan COG (SEMCOG), Atlanta Regional Commission (ARC), Indianapolis MPO, East-West Gateway Council of Government (EWGCOG), Richmond Area MPO, Delaware Valley Regional Planning Commission (DVRPC), North Jersey Transportation Planning Authority (NJTPA), Sacramento Area COG (SACOG), Baltimore Regional Transportation Board (BRTB), Denver Regional COG (DRCOG), Southern California Association of Governments (SCAG), Mid-America Regional Council (MARC). Some travel demand model results are not used due to a lack of access to software or data inconsistency issues but we thank the MPOs who shared them with us including North Central Texas COG (NCTCOG), METROPLAN Orlando, Southeastern Wisconsin Regional Planning Commission (SEWRPC), Memphis Urban Area MPO, state-planning council in State of Rhode Island, Southwestern Pennsylvania Commission (SPC), and Ulster County Transportation Council. 

\section*{Author Contributions}
\textbf{Kaidi Wang:} Conceptualization, Methodology, Software, Validation, Formal analysis, Investigation, Writing Original Draft, Visualization \textbf{Wenwen Zhang:} Conceptualization, Methodology, Software, Validation, Formal analysis, Investigation, Writing Original Draft, Visualization, Supervision

\hfill \break
\textbf{Funding:} This research did not receive any specific grant from funding agencies in the public, commercial, or not-for-profit sectors

\hfill \break
\textbf{Declarations of interest:} none

\hfill\break
\textbf{Color to Print Figure:} please print the figures using color for online distribution

\bibliographystyle{model2-names}\biboptions{authoryear}
\bibliography{trb_template}

\newpage
\section*{Appendix}
\begin{table}[H]
    \caption{Description of Used Travel Demand Models}
    \label{tab: travel demand model}
    \resizebox{\linewidth}{!}
    {
    \begin{tabular}{p{5cm}p{10.5cm}p{1.5cm}p{2.5cm}}
    \toprule
    Metropolitan Area & MPO/Region & Model Type & Model Baseline Year \\
    \midrule
    Virginia Beach, VA & Hampton Roads Transportation Planning Organization (HRTPO) & 4 step & 2018 \\
    San Jose, CA & Metropolitan Transportation Commission (MTC) & ABM & 2015 \\
    Washington, DC--VA--MD & National Capital Region Transportation Planning Board (TPB) & 4 step & 2017 \\
    Salt Lake City, UT & Wasatch Front Regional Council & 4 step & 2015 \\
    Minneapolis, MN--WI & Metropolitan Council & 4 step & 2015 \\
    Chicago, IL--IN & Chicago Metropolitan Agency for Planning (CMAP) & 4 step & 2015 \\
    Phoenix, AZ & Maricopa Association of Governments (MAG)   & 4 step  & 2018 \\
    Las Vegas, NV & Regional Transportation Commission of Southern Nevada (RTC)   & 4 step & 2015 \\
    Seattle, WA & Puget Sound Regional Council (PSRC)   & ABM & 2014 \\
    Detroit, MI & Southeast Michigan COG (SEMCOG)   & 4 step & 2015 \\
    Atlanta, GA & Atlanta Regional Commission (ARC)   & ABM & 2015 \\
    Indianapolis, IN & Indianapolis MPO & 4 step & 2016 \\
    St. Louis, MO--IL & East-West Gateway Council of Government (EWGCOG)   & 4 step & 2015 \\
    Richmond, VA & Richmond Area MPO   & 4 step & 2012 \\
    Philadelphia, PA--NJ--DE--MD & Delaware Valley Regional Planning Commission (DVRPC)   & 4 step & 2015 \\
    New York, NY--NJ--CT & North Jersey Transportation Planning Authority (NJTPA)   & 4 step & 2017 \\
    Sacramento, CA & Sacramento Area COG (SACOG) & ABM & 2012 \\
    Baltimore, MD & Baltimore Regional Transportation Board (BRTB) & 4 step & 2012 \\
    Denver, CO & Denver Regional COG (DRCOG) & ABM & 2015 \\
    San Bernardino, CA & Southern California Association of Governments (SCAG) & 4 step & 2012 \\
    Kansas City, MO--KS & Mid-America Regional Council (MARC)   & 4 step & 2015 \\ 
    \bottomrule
    \end{tabular}
    }
\end{table}
\begin{figure}[H]
    \centering
    \includegraphics[width = \linewidth]{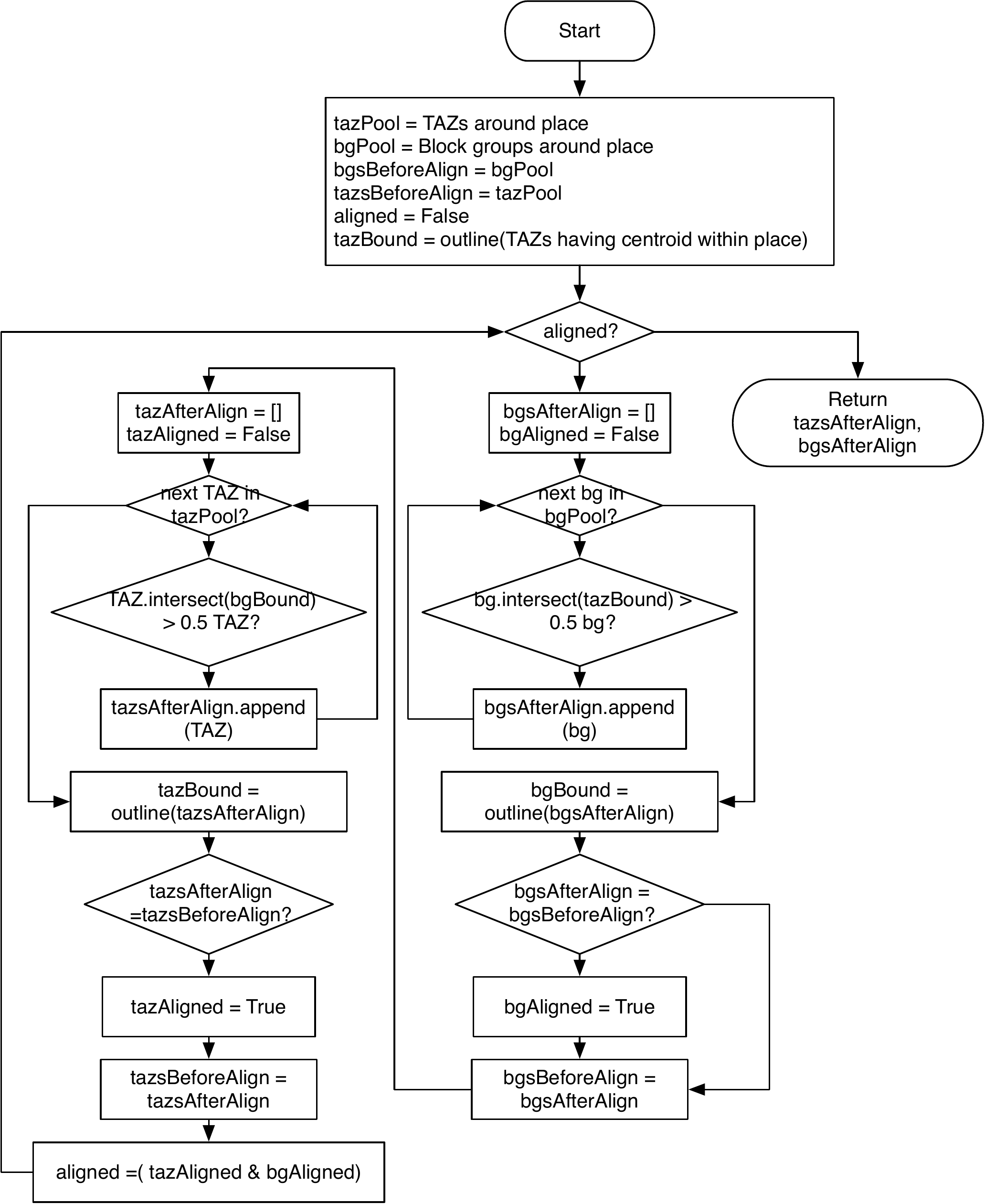}
    \caption{Algorithm to Align the Outlines of TAZs, Census Block Groups, and City Boundaries}
    \label{fig: align}
\end{figure}

\begin{figure}[H]
    \captionsetup[subfigure]{labelformat=empty}
    \begin{subfigure}[b]{0.22\textwidth}
    \centering
    \includegraphics[width=\linewidth]{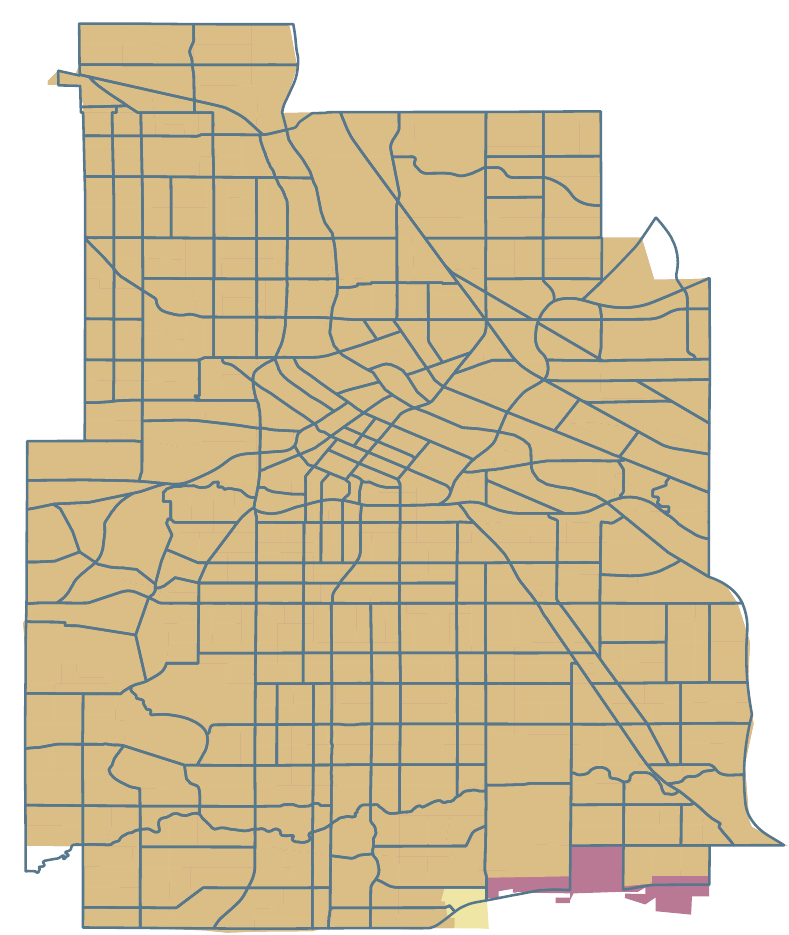}
    \caption{Minneapolis}
    \end{subfigure}
    \hfill
    \begin{subfigure}[b]{0.28\textwidth}
    \centering
    \includegraphics[width = \linewidth]{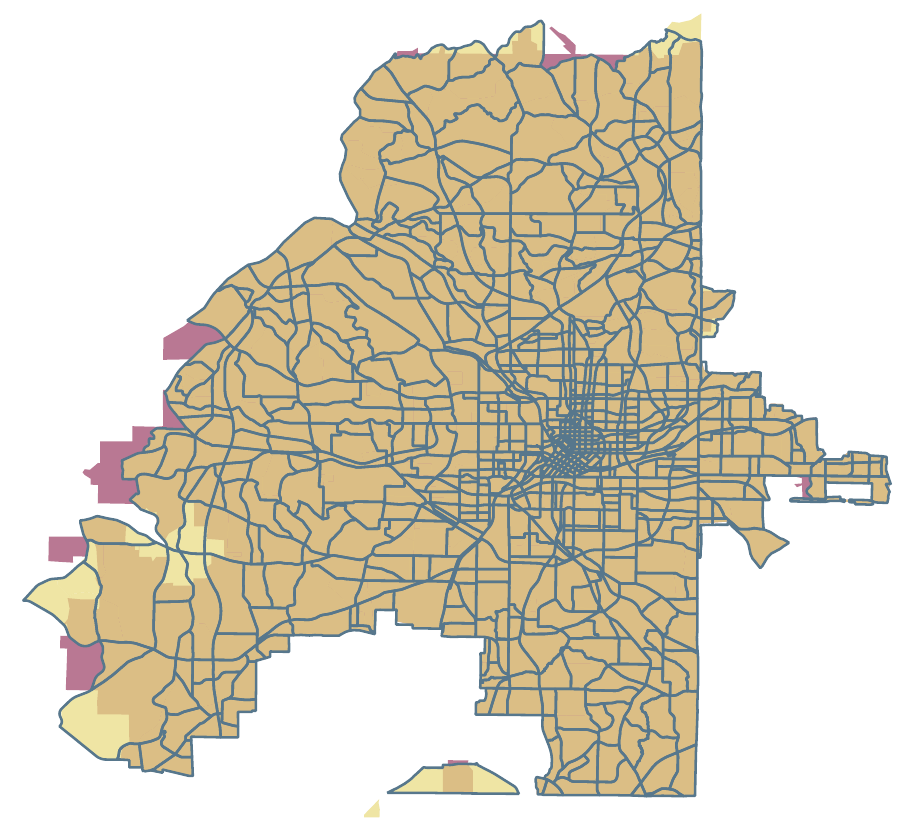}
    \caption{Atlanta}
    \end{subfigure}
    \hfill
    \begin{subfigure}[b]{0.24\textwidth}
    \centering
    \includegraphics[width=\linewidth]{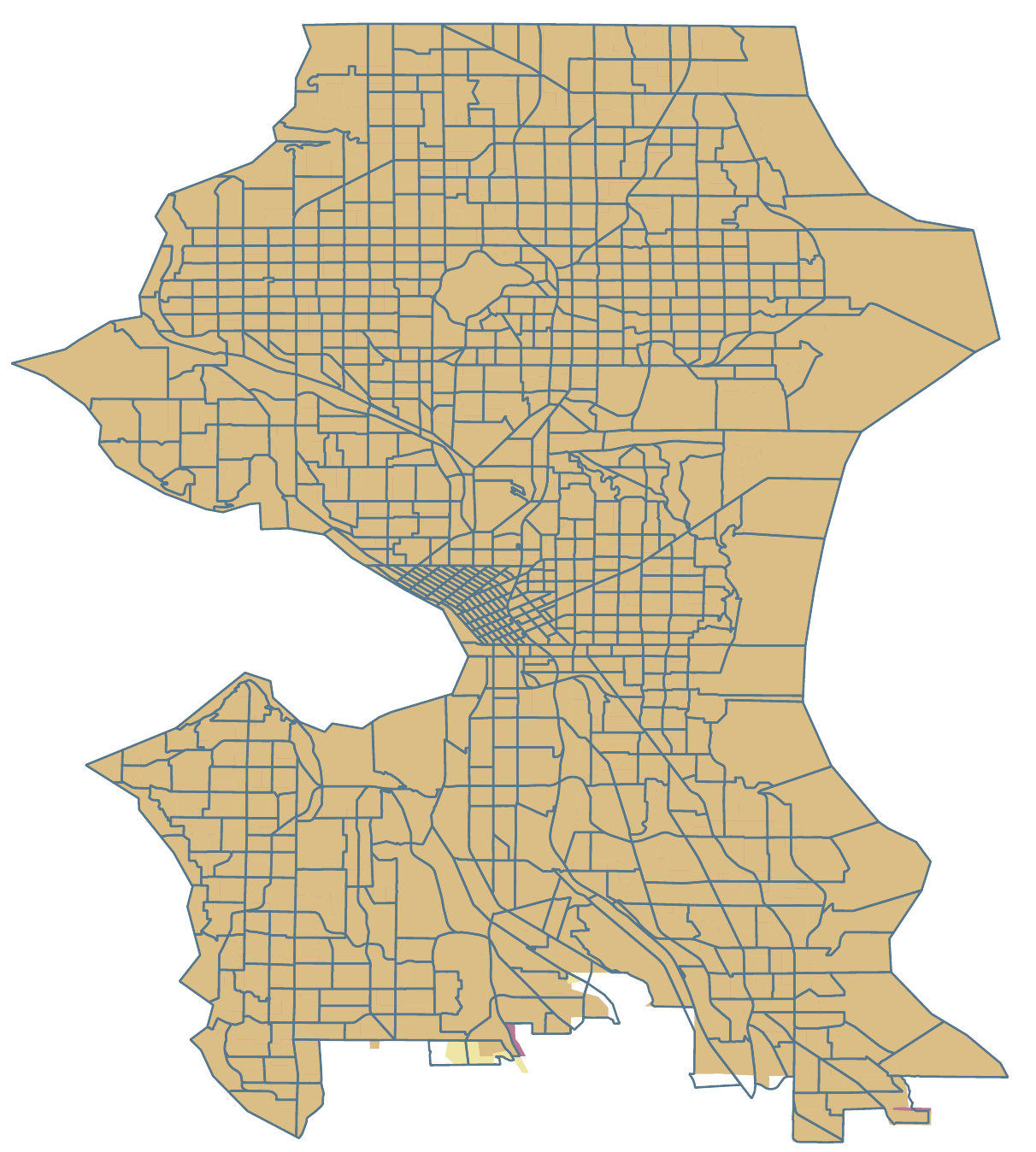}
    \caption{Seattle}
    \end{subfigure}
    \hfill
    \begin{subfigure}[b]{0.23\textwidth}
    \centering
    \includegraphics[height = 4cm]{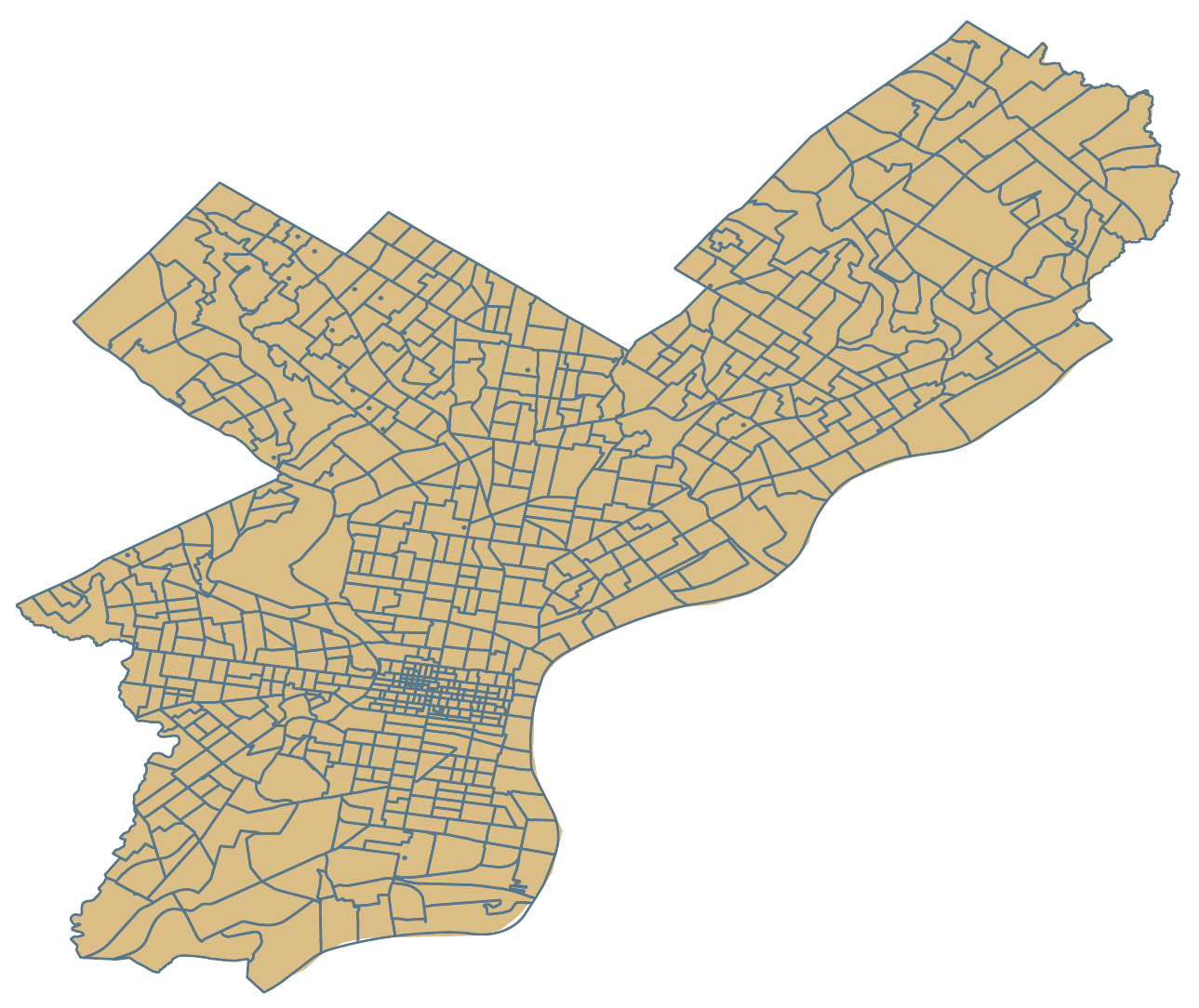}
    \caption{Philadelphia}
    \end{subfigure}
    \centering
    \begin{subfigure}[ht]{0.4\textwidth}
    \centering
    \includegraphics[width = \linewidth]{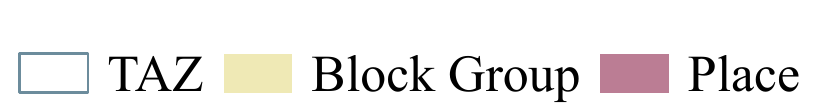}
    \end{subfigure}
    \caption{Examples of Aligned Boundaries for Selected Cities}
    \label{fig: boundary}
\end{figure}

\begin{figure}[H]
    \centering
    \begin{subfigure}[h]{0.32\textwidth}
    \centering
    \includegraphics[width=\linewidth]{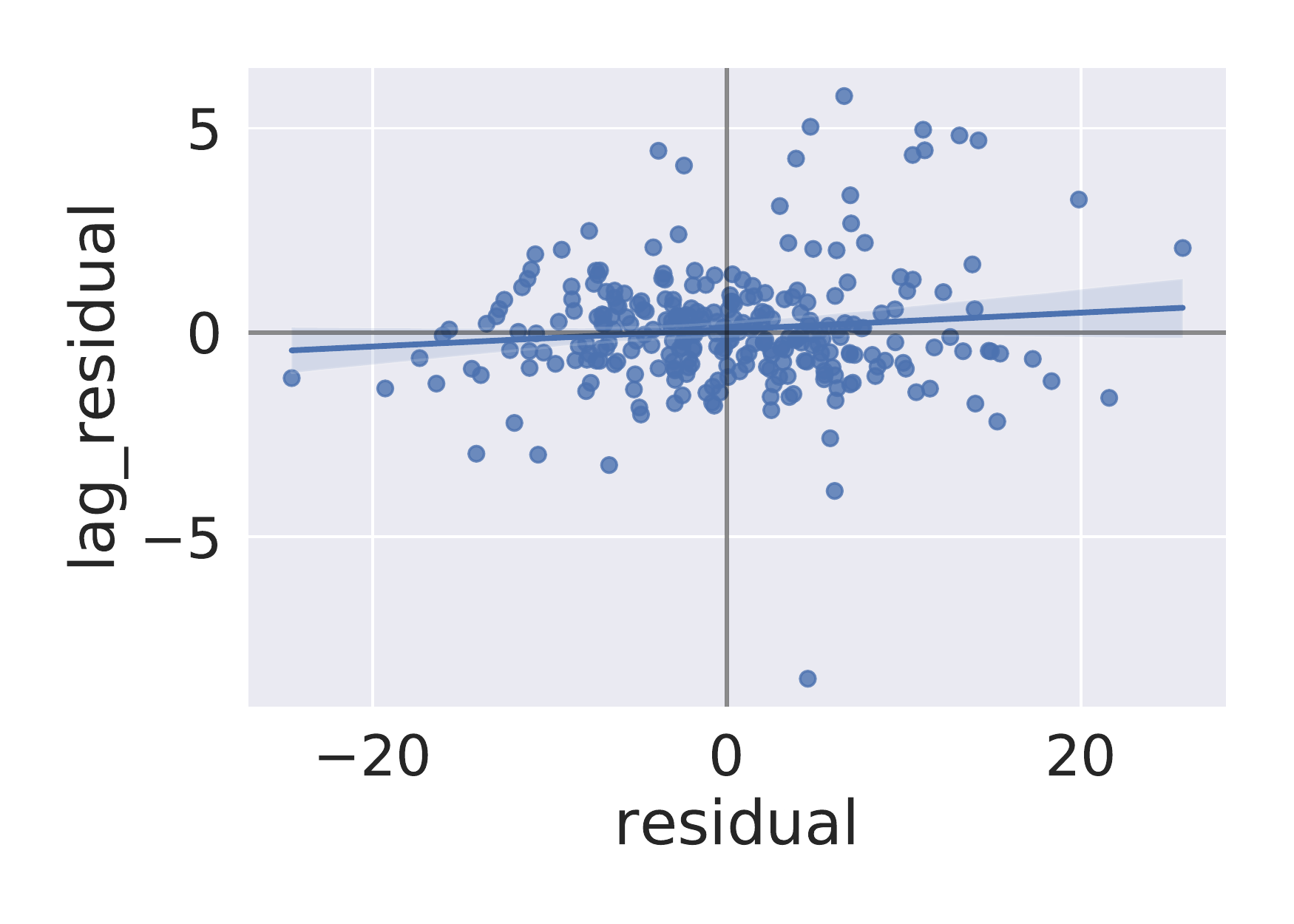}
    \caption{Trips per Vehicle}
    \end{subfigure}
    \begin{subfigure}[h]{0.32\textwidth}
    \centering
    \includegraphics[width=\linewidth]{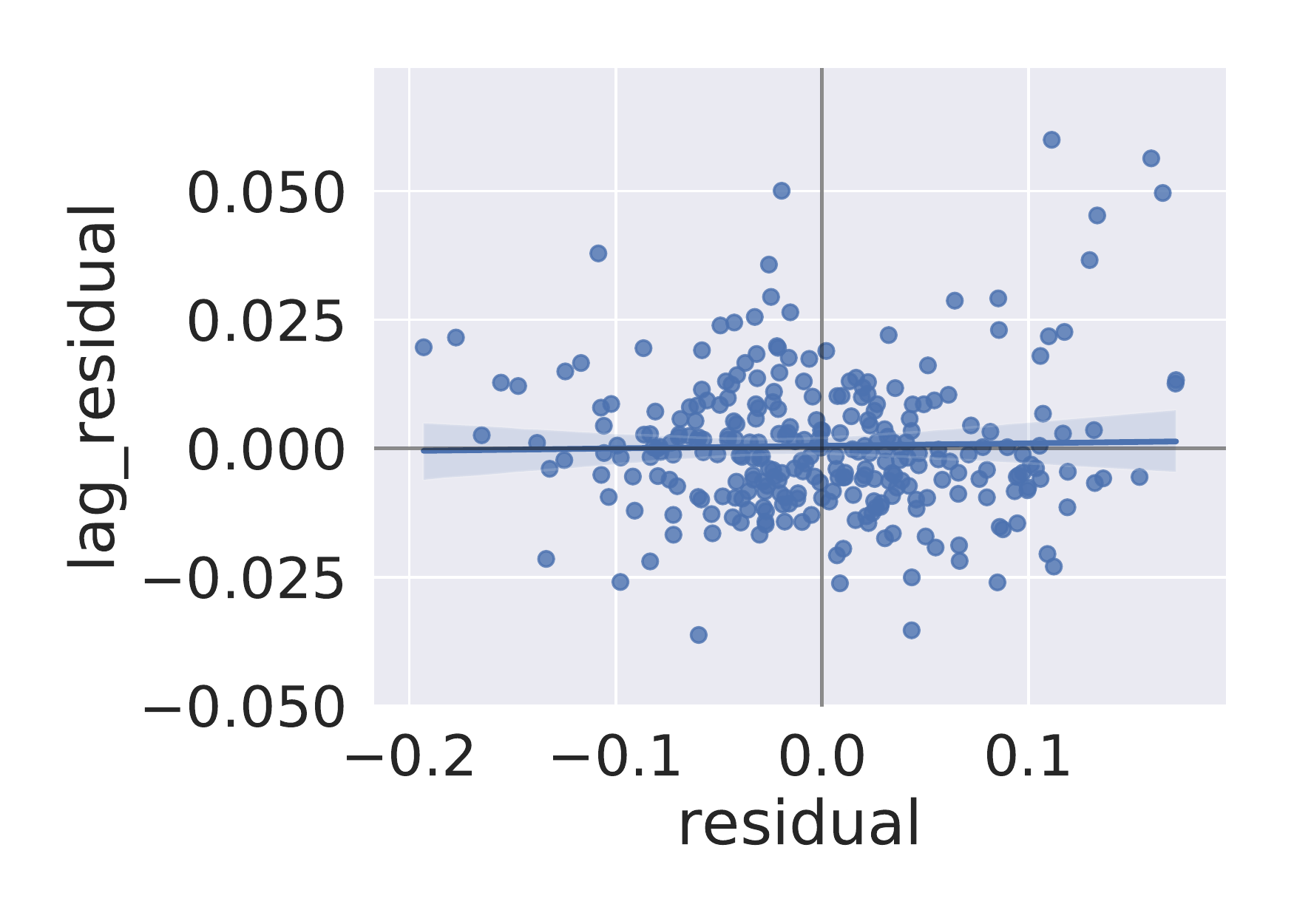}
    \caption{Percent of Pooled Trips}
    \end{subfigure}
    \begin{subfigure}[h]{0.32\textwidth}
    \centering
    \includegraphics[width=\linewidth]{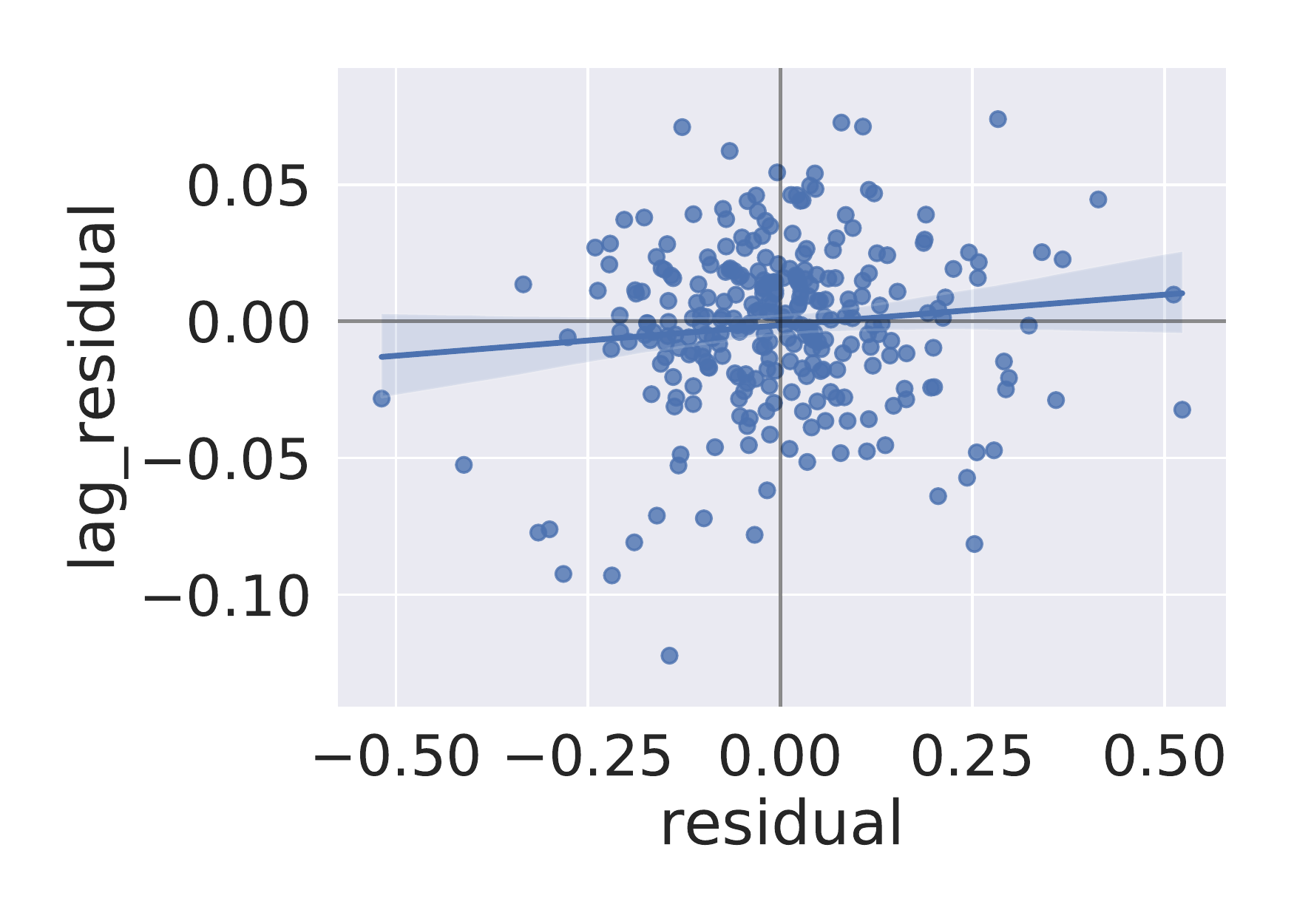}
    \caption{Percent of Extra VMT}
    \end{subfigure}
    \caption{\color{Red}Moran Scatter Plot for Each Model}
    \label{fig:moran}
\end{figure}

\begin{table}[H]
    \centering
    \caption{Model Estimates of the Dummy Variables}
    \label{tab:dummy}\color{Red}
    \begin{tabular}{llll}
    \toprule
        MPO & ServedTrips/SAV & \%pooledTrips & Log(\%extraVMT) \\ \midrule
ARC & -31.666\textsuperscript{***} & 0.042\textsuperscript{*} & 0.298\textsuperscript{***}\\
BRTB & -32.572\textsuperscript{***} & -0.067\textsuperscript{**} & -\\
CMAP & -27.804\textsuperscript{***} & 0.242\textsuperscript{***} & -\\
DRCOG & -5.656\textsuperscript{*} & - & -\\
DVRPC & -20.901\textsuperscript{**} & 0.417\textsuperscript{***} & -0.462\textsuperscript{***}\\
EWGCOG & -15.690\textsuperscript{***} & - & -\\
HRTPO & 6.400\textsuperscript{*} & - & -\\
INDYMPO & - & 0.369\textsuperscript{***} & -0.407\textsuperscript{***}\\
MAG & 4.896\textsuperscript{*} & - & 0.103\textsuperscript{**}\\
MARC & -24.111\textsuperscript{***} & 0.143\textsuperscript{***} & -0.130\textsuperscript{**}\\
MC & -11.141\textsuperscript{***} & -0.090\textsuperscript{***} & -\\
MTC & -14.745\textsuperscript{***} & -0.056\textsuperscript{**} & 0.249\textsuperscript{***}\\
NJTPA & -33.189\textsuperscript{***} & 0.209\textsuperscript{***} & -\\
PSRC & -19.969\textsuperscript{***} & 0.034 & -\\
RTC & -11.254\textsuperscript{***} & -0.101\textsuperscript{***} & -\\
SACOG & - & -0.035 & -\\
SCAG & - & -0.132\textsuperscript{***} & 0.214\textsuperscript{***}\\
SEMCOG & -23.881\textsuperscript{***} & - & -\\
TPB & 11.705\textsuperscript{***} & - & 0.266\textsuperscript{***}\\
Wasatch & - & - & -0.095\textsuperscript{*}\\ \\ \bottomrule
    \end{tabular}
    \begin{tablenotes}
      \small
      \item
      \textsuperscript{***}$p<0.001$, \textsuperscript{**}$p<0.01$, \textsuperscript{*}$p<0.05$.
    \end{tablenotes}
\end{table}

\newgeometry{left=2cm}
\end{singlespace}
\end{document}